\documentclass[10pt,journal,compsoc]{IEEEtran}
\ifCLASSOPTIONcompsoc
  \usepackage[nocompress]{cite}
\else
  \usepackage{cite}
\fi

\usepackage{booktabs}
\usepackage{graphicx}
\usepackage{subcaption}
\usepackage{amssymb}
\usepackage{amsthm}
\usepackage{amsmath}
\usepackage{soul}
\usepackage{amsfonts}
\usepackage{multirow}
\usepackage[ruled]{algorithm}
\usepackage[noend]{algorithmic}
\usepackage{epstopdf}
\usepackage{lipsum}

\newlength{\figwidths}
\newlength{\figwidthd}
\newlength{\expwidths}
\newlength{\expwidthd}
\newcommand{\npc}{\textsc{np}-complete}

\newcommand{\cnpc}{\textsc{coNP}-complete}
\newcommand{\rdf}{\textsc{RDF}~}

\newcommand\blfootnote[1]{%
  \begingroup
  \renewcommand\thefootnote{}\footnote{#1}%
  \addtocounter{footnote}{-1}%
  \endgroup
}

\makeatletter
\renewenvironment{proof}[1][\proofname]{\par
	\pushQED{\qed}%
	\normalfont \partopsep=\z@skip \topsep=\z@skip
	\trivlist
	\item[\hskip\labelsep
	\itshape
	#1\@addpunct{.}]\ignorespaces
}{%
	\popQED\endtrivlist\@endpefalse
}
\makeatother

\newtheorem{definition}{Definition}
\newtheorem{example}{Example}
\newtheorem{problem}{Problem}
\newtheorem{theorem}{Theorem}
\newtheorem{lemma}[theorem]{Lemma}
\newtheorem{proposition}[theorem]{Proposition}
\ifCLASSINFOpdf
\else
\fi
\begin{document}
%
\title{Dynamic Relation Repairing for Knowledge Enhancement}
%
%
%
%
\author{Rui Kang, Hongzhi Wang*\thanks{Pengcheng Lab, Harbin Institute of Technology, email: wangzh@hit.edu.cn}}

%
%

\markboth{Journal of \LaTeX\ Class Files,~Vol.~14, No.~8, August~2015}%
{Shell \MakeLowercase{\textit{et al.}}: Bare Demo of IEEEtran.cls for Computer Society Journals}
%



\IEEEtitleabstractindextext{%
\begin{abstract}
Dynamic relation repair aims to efficiently validate and repair the instances for knowledge graph enhancement (KGE), where KGE captures missing relations from unstructured data and leads to noisy facts to the knowledge graph.
With the prosperity of unstructured data, an online approach is asked to clean the new \textsc{RDF} tuples before adding them to the knowledge base.
To clean the noisy \textsc{RDF} tuples, graph constraint processing is a common but intractable approach.
Plus, when adding new tuples to the knowledge graph, new graph patterns would be created, whereas the explicit discovery of graph constraints is also intractable.
Therefore, although the dynamic relation repair has an unfortunate hardness, it is a necessary approach for enhancing knowledge graphs effectively under the fast-growing unstructured data.
Motivated by this, we establish a dynamic repairing and enhancing structure to analyze its hardness on basic operations.
To ensure dynamic repair and validation, we introduce implicit graph constraints, approximate graph matching, and linkage prediction based on localized graph patterns.
To validate and repair the \textsc{RDF} tuples efficiently, we further study the cold start problems for graph constraint processing.
Experimental results on real datasets demonstrate that our proposed approach can capture and repair instances  with wrong relation labels dynamically and effectively.
\end{abstract}

}

\maketitle

\IEEEdisplaynontitleabstractindextext

%
\IEEEpeerreviewmaketitle

\section{Introduction}
\label{sec:intro}
\blfootnote{This paper was supported by NSFC grant (U1866602,71773025). The National Key Research and Development Program of China (2020YFB1006104).}
Knowledge graph has been the backbone of many information systems for searching the dependency relationship of entities~\cite{DBLP:journals/corr/abs-2004-11861,DBLP:journals/corr/abs-1903-02419,Zheng2018}.
They are constructed in large size to cover more information to provide a better performance, such as \textsc{YAGO}~\cite{DBLP:conf/www/HoffartSBLMW11}, Freebase~\cite{Bollacker2008}, and WikiData.
Knowledge acquisition applies relation extraction or information extraction on three main sources, unstructured data(plain text), \textsc{HTML} schemas, and human-annotated pages \cite{Dong2014}.
Unlike other sources of knowledge graphs being seen plateaued~\cite{DBLP:conf/wikis/SuhCCP09}, knowledge acquisition from plain text is always used to build large and scalable knowledge bases \cite{Dong2014}.
Both knowledge acquisition tools and knowledge databases are seeking online obtaining and enhancing approaches as unstructured data has been seen growing at a rapid speed~\cite{DBLP:journals/jiis/ProtaziukLB16}.

In this paper, we construct \rdf knowledge database from scalable unstructured data ~\cite{Zeng2015,Han2018,DBLP:journals/corr/abs-1902-00756} with each tuple in the form of $\langle \mathit{head}, \mathit{relation}, \mathit{tail} \rangle$.
\rdf~ tuples composed by \textsc{Top-1} relations will be the most fundamental part of the knowledge graph.
However, according to the \textsc{Top-1} statistics provided by recent works~\cite{DBLP:journals/corr/abs-1906-00687,DBLP:journals/corr/abs-1911-00219} and \textsc{HIT@N}~\cite{Han2018,Riedel2013,DBLP:journals/corr/abs-1902-00756}, knowledge acquisition tools often generate noisy and unreliable facts~\cite{Weikum2010}.
Some facts would even introduce violations to the knowledge graph affecting the applications of knowledge graph.
We use Example \ref{exp:intro} to illustrate the conflicts brought by relation extracting.

\begin{figure}
	\centering
	\includegraphics[width=0.45\textwidth]{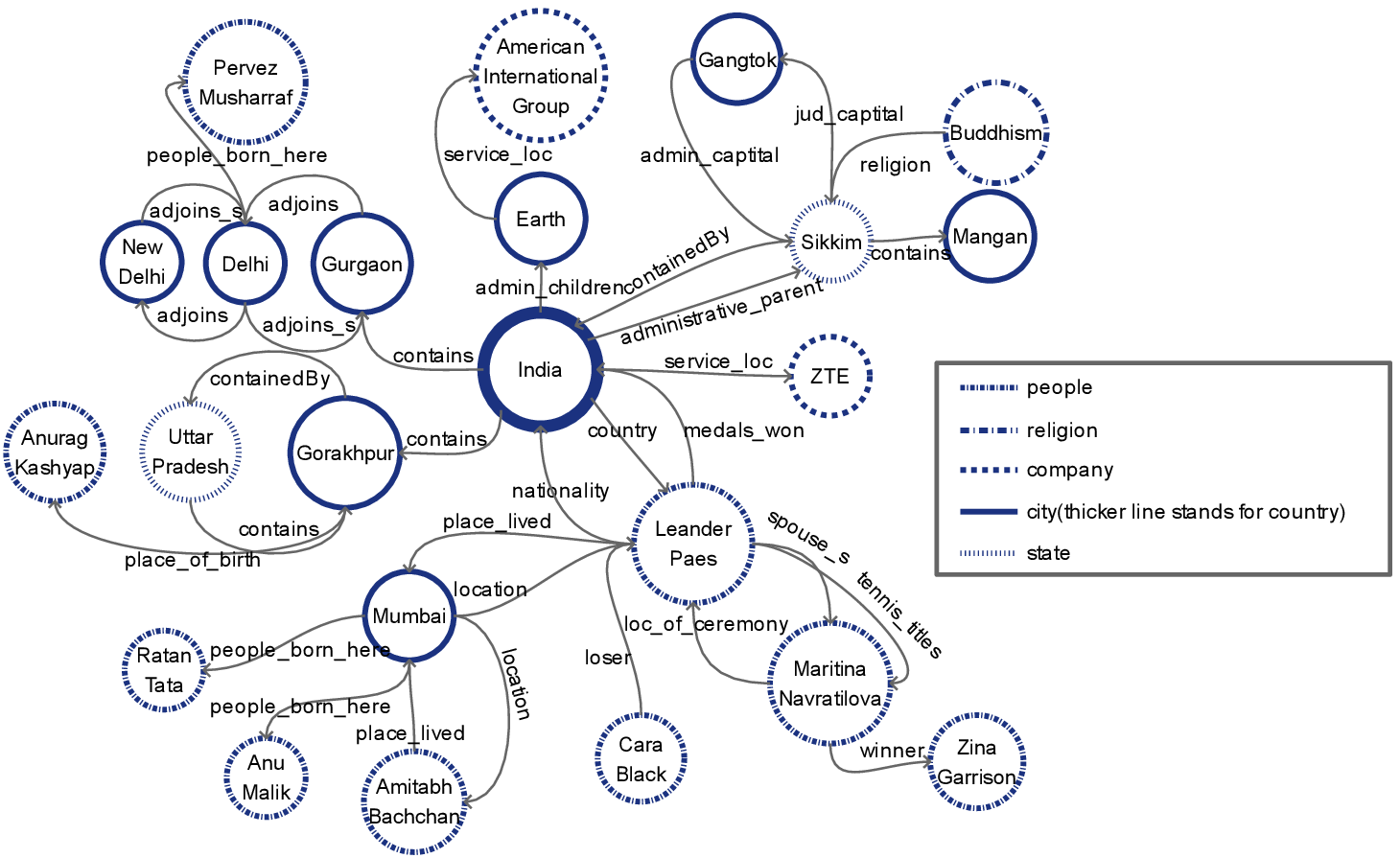}
	\caption{The fragment of a knowledge graph}
	\label{fig:demo}
\end{figure}

\begin{example}
	\label{exp:intro}
	Consider the linkage between ``India'' and ``Leander Paes'' in Figure \ref{fig:demo}.
	Given a sentence $c:$ ``Leander Paes won a bronze medal for India in singles in the 1996 Atlanta Olympic Games.'' and a pair of entities $\langle h,t \rangle = \langle \mathit{Leander\_Paes}, \mathit{India} \rangle$, a knowledge acquisition tool would consider both $c$ and $\langle h,t \rangle$ to generate its prediction on the relation label between $h$ and $t$.
	The prediction is a map between relation labels and their probabilities e.g., $\mathcal{P} = \{(\mathit{contains}, 0.42), (\mathit{medals\_won}, 0.33), ...\}$.
	Under this prediction, we would consider an addition of \textsc{RDF} tuple with \textsc{Top-1} probability relation: $\langle \mathit{Leander\_Paes}, \mathit{contains}, \mathit{India} \rangle$ instead of the true label $\mathit{medal\_won}$.
	The insertion would cause violations in the knowledge graph like Figure \ref{fig:demo}, because relation label-$\mathit{contains}$ should not exist as an edge label from a person to a country, which could be caught by a graph constraint \cite{DBLP:journals/jdiq/Fan19}.
\end{example}

As shown by Example \ref{exp:intro}, new instances by knowledge acquisition without consistency checking may cause consistency and stability issues in the knowledge database.
As knowledge databases confront an online addition with large amount of \rdf tuples acquiring from prosperous unstructured data recently, an online consistency checking is important for knowledge enhancement.

Existing works involving graph relation repair can be classified into two main approaches i.e., graph constraints \cite{DBLP:journals/jdiq/Fan19,DBLP:conf/sigmod/FanHLL18,DBLP:journals/tods/Song011} and linkage prediction through graph representation \cite{Dong2014,DBLP:journals/ijccc/Lin0WWYG18,DBLP:journals/ijswis/PaulheimB14}.
Given a set of graph constraints, rule based graph instance repairing involves the problem of graph validation, which is proved intractable under general case \cite{DBLP:journals/jdiq/Fan19}.
Moreover, to capture new graph patterns in graph constraints from KG, the graph constraint discovery relates to the intractable satisfiability and implication problems \cite{DBLP:journals/jdiq/Fan19}, which also leads to dynamic hardness.
Knowledge graph representation mostly relates to path ranking or neural network, which requires a pretraining process and would fail on both real-time instance updating and cold start problem e.g., a new entity to KG.
In conclusion, the existing works fail on solving the dynamic relation repair problem.
In this paper, we combine the above two kinds of approaches together by dynamically considering potential relation labels based on linkage prediction and graph constraint validation.

As one of our main motivations, we firstly notice that besides applying existing constraints for graph quality, the discovery of newly formed constraints is also important for efficient relation repair.
Moreover, as constraints processing would confront dynamic hardness, simplifying instance validation and constraint discovery is a must.
In this study, we build an algorithmic structure on repairing and enhancing under dynamic situation.
Implicit graph constraint avoids the discovery operation and converts the validation into an optimizable process i.e., subgraph matching.
To advance graph matching in dynamic KG, we propose a graph embedding method based on the label information in subgraphs.
To achieve relation repair on conflict instances, we could naively enumerate possible instances generated by knowledge acquisition and validate them based on graph constraints.
Since linkage prediction is only based on the knowledge graph and can identify impossible relation labels, we use linkage prediction as a supervisor to prune impossible instances.
Our repair model is launched based on the above motivations by optimizing the knowledge acquisition results with linkage prediction and validating possible instances.

\subsection*{Contributions}

Our major contributions in this paper are summarized as follows.

(1) We propose an algorithmic structure over graph constraints for dynamic knowledge graph enhancing as the first structure considering all three operations - validation, repair, and discovery for sustainability and consistency of the knowledge graph.
	
(2) We formalize and devise an optimize-validate model for the relation label repair problem.
In this model, we optimize the knowledge acquisition results with linkage prediction and validate the generated instances with graph constraints.
We propose an evaluation for linkage prediction based on graph matching, which is optimizable.

(3) We introduce implicit graph constraints to eliminate graph constraints discovery operation and convert instance validation into an optimizable process involving graph matching.

(4) We notice the hardness brought by the cold start problem during knowledge enhancement and propose applicable solutions, where two possible situations are discussed including fresh entities and rare relation labels.
	
(5) Finally, experimental results on real datasets show our approach is a dynamic process with an evident improvement on precision and F-score of \textit{Top-1} possible relation.
The performance of relation repair on scalable datasets indicates that our approach can aid both sustainability and consistency of knowledge database.

\section{Problem Statement and Overview}
\label{sec: problemstatement}

In this section, we formalize the problem of repairing relation for knowledge enhancement (Section \ref{sec:bgmProblemDefinition}).
To analyze the dynamic situation under scalable unstructured data source, we propose a general structure for dynamic repairing and enhancing (Section \ref{sec:overviewDynamic}).
As a part of the dynamic structure, the optimize-validate relation repair model is proposed for repairing the violated instances (Section \ref{sec:cvRelationRepair}).
We analyze the hardness we meet for dynamically repairing and enhancing the knowledge graph in Section \ref{sec:overviewDynamic} and Section \ref{sec:cvRelationRepair}

\subsection{Background and Problem Definition}
\label{sec:bgmProblemDefinition}

We analyze the notations and the dynamic relation repair problem in this section.

\setuldepth{Database}
\ul{\textbf{Graph Database}.}
Graph database $\mathcal{G}$ is a collection of \textsc{RDF} tuples $s=\langle h,r,t\rangle \in \mathcal{G}$ and also written as $\mathcal{G} = \{V,E,L\}$, where $V$ indicates the set of nodes and $E$ is the set of directed edges. 
Label function $L$ is a mapper $L: V \cup E \to L_\mathcal{G}$, where $L_\mathcal{G} = L_\mathcal{E} \cup L_\mathcal{V}$ denotes the union of edge and vertice labels, respectively.
We consider $|\mathcal{G}| = |\mathcal{G}.V|$ as the size of $\mathcal{G}$.
Given tuple $s \in \mathcal{G}$, we denote $G(s) \subset \mathcal{G}$ as a subgraph containing $s$. 


\ul{\textbf{Knowledge Acquisition}.}
Given each sentence $c_i$ in corpus $\mathcal{C}$ with a pair of entities $\langle h_i,t_i \rangle$, knowledge acquisition is a classifier on multiple relation labels, which generates a projection between relation labels and their probabilities i.e., $\mathcal{P}_i = \{(r_j, p_j)|p_j = Pr(r_j|\langle h_i,r_i \rangle, c_i), c_i\in\mathcal{C}\}$, where the relation labels are taken from $r \in L_\mathcal{E} \cup \{\mathit{NA}\}$.
\textit{NA} is a negative label indicating that no relation could be detected between $h_i$ and $t_i$ in sentence $c_i \in \mathcal{C}$.
When considering a deletion on tuple $\langle h,r,t \rangle \in \mathcal{G}$, we modify $r$ into \textit{NA}.
$m_i$ holds the projections from $\mathcal{P}_i$ with \textsc{Top}-k probabilities i.e., $m_i \subsetneq \mathcal{P}_i, |m_i|=k$.
Therefore, we obtain a candidate set $\mathcal{M} = \{m_i| c_i \in \mathcal{C}\}$ and an instance set $\mathcal{I} = \{\langle h_i,r_i,t_i \rangle|(r_i, p_i) = \arg \max_{(r_j, p_j) \in m_i} p_j \wedge p_i \geq p_{th}, m_i \in \mathcal{M}\}$.
We use $p_{th}$ as the threshold for the predicting probability of relation labels to become candidates, and we will not add a tuple when $p_{i,j} < p_{th}, \forall r_j \in L_\mathcal{E}$ given sentence $c_i$ and entities $\langle h_i, t_i \rangle$.

\ul{\textbf{Graph Matching}.}
A match $\gamma$ considers the isomorphism between two subgraphs $G_1$ and $G_2$, i.e., $\gamma$ is a bijection between two vertex sets $G_1.V$ and $G_2.V$ with $\forall u \in G_1.V, \gamma(u) \in G_2.V$ and $\forall v \in G_2.V, \gamma^{-1}(v)\in G_1.V$.
The match $\gamma$ preserves the equivalence on labels and vertice neighborhood i.e., $\forall u \in G_1.V, G_1.L(u) = G_2.L(\gamma(u))$ and $\forall e(u_1,u_2)\in G_1.E, \gamma(e) = e(\gamma(u_1), \gamma(u_2)) \in G_2.E \wedge G_1.L(e) = G_2.L(\gamma(e))$.

\ul{\textbf{Graph Constraints}.}
Before constructing a graph database, a set of constraints $\Sigma$ is predefined for database consistency.
$\varphi$, one of the elements in $\Sigma$, can be written in the form as $\varphi = Q[\overline{x}](X \Rightarrow Y)$\cite{DBLP:journals/jdiq/Fan19}, where $Q[\overline{x}]$ is a connected graph pattern, $X,Y$ are sets of literals and $X \to Y$ is the logistic expression in a graph constraint.
Variables involved in $X,Y$ are listed in $\overline{x}$.
The size of $Q[\overline{x}]$ is equal to the number of vertices i.e., $|Q[\overline{x}]| = |\overline{x}|$.
We use $G'\vDash X$ or $G'\vDash Y$ when the related variables satisfy the literals in $X$ or $Y$, respectively.
$G'\vDash\varphi$ is to express that $G'$ is a match of $Q[\overline{x}]$ and when $G'\vDash X$, $G'\vDash Y$ should also be satisfied $(G'\vDash X\to Y)$.
We say $\mathcal{G}\vDash \varphi$ only when for each matched subgraph $G'$ of $\varphi.Q$, $G' \vDash X \to Y$.
Given graph $\mathcal{G}$ and a set of graph constraints $\Sigma$, $\mathcal{G} \vDash \Sigma$ is satisfied only when each constraint is satisfied ($\mathcal{G} \vDash \varphi, \forall \varphi \in \Sigma$).

\ul{\textbf{Relation Label Repair}.}
Relation repair aims to find a repair $\mathcal{I}'$ for $\mathcal{I}$ w.r.t. the graph database $\mathcal{G}$ and a set of graph constraints $\Sigma$  over $\mathcal{G}$ i.e., $\mathcal{I}' \cup \mathcal{G} \vDash \Sigma$.
The repair modifies a subset of $\mathcal{I}$ with each \textsc{RDF} tuple choosing a new relation label from $L_\mathcal{E} \cup \{NA\}$.

\ul{\textbf{Knowledge Enhancement}.}
Knowledge enhancement is a process of generating and adding validated instance set $\mathcal{I}$ into $\mathcal{G}$, written as $\mathcal{G} \leftarrow \mathcal{G} \cup \mathcal{I}$ s.t. $\mathcal{G} \cup \mathcal{I} \vDash \Sigma$.
When $\mathcal{G} \cup \mathcal{I}$ violates the graph constraints $\Sigma$ defined over $\mathcal{G}$, a repair $\mathcal{I}'$ based on $\mathcal{I}$ is required to make $\mathcal{G}\cup\mathcal{I}' \vDash \Sigma$ satisfied.


\begin{problem}
\ul{\textbf{Dynamic Relation Repair}.} 
Given the stream $\mathbb{F}$ of noisy \textsc{RDF} tuples based on corpus stream $\mathcal{C}$, dynamic relation repair aims to find a clean \textsc{RDF} stream $\mathbb{F}'$ as real-time KG enhancement without causing violations to database $\mathcal{G}$.
\end{problem}

Dynamic relation repair could advance the applicability of multiple approaches such as knowledge enhancement and dynamic event graph repairing \cite{DBLP:conf/icde/0001SLZP15}, which involves the violation detection and repair over noisy graph data stream.
The naive approach to dynamic relation repair is to find a violation-free enhancement at each time slot by checking and repairing the enhanced graph $\mathcal{G}(T + \Delta T) \leftarrow \mathcal{G}(T) \cup \mathbb{F}(\Delta T)$ with a set of predefined graph constraints $\Sigma$.
Unfortunately, the naive approach is highly intractable under current graph quality approaches.
The validation problem aims to check whether the graph database satisfies a set of graph constraints i.e., $\mathcal{G} \vDash \Sigma$, which is proved to be \cnpc~under graph functional dependencies (GFDs) \cite{DBLP:journals/jdiq/Fan19}.
To capture new graph patterns generated by incremental graphs, the update of graph constraints involves the satisfiability and implication problems.
Among them, the satisfiability problem finds the support of $\forall \varphi \in \Sigma$ in graph $\mathcal{G}$ and detects conflict among $\varphi \in \Sigma$, which is intractable (\cnpc) even under GFDs with constant literals; the implication problem prunes unnecessary graph quality rules, which is \npc~under general GFDs.

\subsection{Overview of Dynamic Relation Repair}
\label{sec:overviewDynamic}

We introduce this section to analyze the hardness of processing repair and enhancement under scalable unstructured data.
As knowledge acquisition can produce noisy results and introduce violations to the graph database, we consider a dynamic relation repair over knowledge acquisition results.

While merging new \textsc{RDF} tuples into the graph database $\mathcal{G}$, the structural information of $\mathcal{G}$ is updated and so as the structural feature embedded in $\mathcal{G}$.
Moreover, due to the hardness of graph constraint discovery, limited constraints can be defined or inherited when initializing the knowledge database, and most of the constraints are needed to be discovered after the database is effectively enhanced \cite{Bollacker2008,DBLP:conf/sigmod/FanHLL18,DBLP:conf/www/HoffartSBLMW11}.
Therefore, we need to take a discovery approach for possible new constraints formed during the enhancing process.
Dynamic repair and enhancement should include a circle formed by four operations i.e. acquisition-validate/repair-enhance-discovery.
Motivated by this, dynamic repairing algorithms for knowledge enhancement need to follow a common structure in Algorithm \ref{algorithm:dynamicStructure}.

\begin{algorithm}[h]
	\small
	\caption{Dynamic repair and enhancing structure} %
	\label{algorithm:dynamicStructure} %
	\begin{algorithmic}[1]
		\REQUIRE Initial graph $\mathcal{G}_0$, Graph constraints $\Sigma_0$ and Corpus stream $\mathcal{C}$.
		\ENSURE Enhanced graph after $T$ iterations $\mathcal{G}_T$ and Constraints $\Sigma_T$.
		\STATE $\mathcal{G} := \mathcal{G}_0$.
		\STATE $\Sigma := \Sigma_0$
		\FOR{streaming slice index $i$}
		\STATE Fetch current corpus $\mathcal{C}_i$.
		\STATE Generate prediction $\mathcal{M}_i := \mathit{KnowledgeAcquire}(\mathcal{C}_i)$.
		\STATE Initialize instance $\mathcal{I}_i$ with $\mathcal{M}_i$
		\IF{Validation($\mathcal{G}\cup\mathcal{I}_i \not\vDash \Sigma$)}
		\STATE Find a repair $\mathcal{I}_i = \mathit{Repair}(\mathcal{I}_i, \mathcal{G}, \Sigma)$.
		\ENDIF
		\STATE Update graph $\mathcal{G} = \mathcal{G}\cup\mathcal{I}_i$.
		\STATE Constraint update $\Sigma = \mathit{Discovery}(\mathcal{G}, \Sigma)$.
		\STATE Send back current $\mathcal{G}$ and $\Sigma$ if asked.
		\ENDFOR
		\RETURN $\mathcal{G}, \Sigma$
	\end{algorithmic}
\end{algorithm}

Based on Algorithm \ref{algorithm:dynamicStructure}, we use $T_{\mathit{valid}}(\mathcal{I}_i)$, $T_{\mathit{repair}}(\mathcal{I}_i)$, $T_{\mathit{discovery}}(\mathcal{I}_i)$ to represent the time consumption of instance validation, instance repair and graph constraints discovery respectively in one iteration $i$.
Suppose we could obtain the label-probability set $\mathcal{P}$ in $O(1)$ time for each sentence in $\mathcal{C}_i$ with the knowledge acquisition tools, the amortized time for enhancing process would be decided by validation/repair-discovery process i.e., $\tilde{O}(\frac{T_{\mathit{valid}}(\mathcal{I}_i)+T_{\mathit{repair}}(\mathcal{I}_i)+T_{\mathit{discovery}}(\mathcal{I}_i)}{|\mathcal{I}_i|})$.
We further assume that the maximum graph pattern size of each rule in $\Sigma$ is restricted by a constant $q_m$ i.e., $|\varphi.Q[\overline{x}]|\leq q_m, \forall \varphi \in \Sigma$, which leads to an upper bound of the complexity of constraint discovery i.e., $T_{\mathit{discovery}}(\mathcal{I}_i) = O(\frac{|\mathcal{G}_i\cup\mathcal{I}_i|^{q_m}}{n})$ given $n$ as the parallelism degree \cite{DBLP:conf/sigmod/FanHLL18}.
The validation process is \cnpc~ w.r.t. the size of $\Sigma$ and $\mathcal{G} \cup \mathcal{I}_i$ \cite{DBLP:journals/tods/FanL19}.
The complexity of instance repairing depends on the strategy we take, but it should include a validation process to check the consistency of the repaired instance, which is \cnpc.
The repairing strategy is analyzed in detail in section \ref{sec:cvRelationRepair}.
We could observe that the operations involving in Algorithm \ref{algorithm:dynamicStructure} meet dynamic hardness.
To process dynamically, the amortized complexity should be $\tilde{O}(1)$.

\subsection{Optimize-Validate Relation Repair Model}
\label{sec:cvRelationRepair}


In this work, each \textsc{RDF} tuple $\langle h, r, t \rangle$ in the instance $\mathcal{I}$ is optimized by the joint probability (supervised by both graph structural information in $\mathcal{G}$ and the corpus $\mathcal{C}$) and validated by the graph constraints $\mathcal{G} \cup \mathcal{I} \vDash \Sigma$.
This is because (1) the knowledge graph could provide structural information such as the interconnection among entities, which would help to prune impossible relation labels and optimize the prediction of the linkages between a pair of tuples; (2) constraints can aid the detection of illegal relation labels with graph patterns. 
Therefore, to repair the relation labels for general knowledge acquisition tools, we use the joint probability $Pr(\mathcal{I}|\mathcal{C,G})$ to consider the results from both corpus extraction and graph structure supervision.

According to the description of knowledge acquisition formalized in Section \ref{sec:bgmProblemDefinition}, the initial instance $\mathcal{I}$ is further writen as $\mathcal{I} = \arg \max_{\mathcal{I}_x} \mathit{Pr}(\mathcal{I}_x|\mathcal{C})$ with $Pr(\mathcal{I}_x|\mathcal{C}) = \prod_{\langle h_i,r_i,t_i \rangle \in \mathcal{I}_x, c_i \in \mathcal{C}} Pr(r_i|\langle h_i,t_i \rangle, c_i)$.
To repair the noisy instance $\mathcal{I}$ by deciding the relation label of each \textsc{RDF} tuple, we consider both the structural information embedded in graph $\mathcal{G}$ and the instance prediction $Pr(\mathcal{I}|\mathcal{C})$ by the joint probability $Pr(\mathcal{I}|\mathcal{C}, \mathcal{G}) = \mathit{Pr}(\mathcal{I}|\mathcal{C})*\mathit{Pr}(\mathcal{I}|\mathcal{G})$.

Linkage prediction aims to find possible relations between a pair of entities based on a knowledge graph, which is studied in Section \ref{sec:link_prediction}.
We could use linkage prediction to generate the probability of each relation label between $\langle h,t \rangle$ i.e., the conditional probability $\mathit{Pr}(\mathcal{I}|\mathcal{G})$.
To find a valid instance $\mathcal{I}'$ and maximize the joint prediction $\mathit{Pr}(\mathcal{I}|\mathcal{C,G})$, an optimize-validate model is proposed considering a set of constraints $\Sigma$ defined over $\mathcal{G}$.
Therefore, we have $\mathcal{I}' = \arg\max_{\mathcal{I}' \vDash \Sigma} Pr(\mathcal{I}'|\mathcal{C},\mathcal{G})$ as the repair of the initial instance $\mathcal{I}$.

Given $\mathcal{M}_{\mathit{lp}}, \mathcal{M}_{\mathit{ka}}$ as the result of linkage prediction (LP) and knowledge acquisition (KA), we could find the optimized answer $\mathcal{I}'$ through instance validations on each instance candidate $\mathcal{I}'$.
Consider parameter $k$ as \textsc{Top}-k probabilities in $\mathcal{M}_{\mathit{lp}}, \mathcal{M}_{\mathit{ka}}$, the instance $\mathcal{I}'$ is a combination of tuples with each tuple having $k$ choices on the relation label.
Observe this, regarding $T_{\mathit{lp}}$ as the time consumption of linkage prediction for instance $\mathcal{I}$, the instance repairing through optimization and validation terminates under amortized time $\tilde{O}(\frac{k^{|\mathcal{I'}|}*T_{\mathit{valid}}(\mathcal{I'})+T_{\mathit{lp}}}{|\mathcal{I}'|})$, which confronts a dynamic hardness.

\begin{example}
\label{exp:optimize-validate}
We extend the Example \ref{exp:intro} and use $\mathcal{M}_{\mathit{ka}}$ for the knowledge acquisition result on the corpus $\mathcal{C}$.
The linkage prediction result $\mathcal{M}_{\mathit{ka}}$ shares the same data schema as $\mathcal{M}_\mathit{ka}$ e.g., $\mathcal{M}_{\mathit{lp}} = \{\{(\mathit{medals\_won}, 0.72),(\mathit{contains}, 0.31), ...\}\}$, which indicates the probability on relation labels between a pair of entities.
For each item in $\mathcal{M}_{\mathit{ka}}$ and $\mathcal{M}_{\mathit{lp}}$, we perform a join operation to compute the joint prediction.
We have $\mathcal{M}_{\mathit{res}} = \{\{$ $(\mathit{medals\_won},$ $0.25),$ $(\mathit{contains}, 0.12)...$ $\}\}$.
Two repair candidates are visible $\mathcal{I}'_1=\{\langle$ $\mathit{Leander\_Paes},$ $\mathit{contains},$ $\mathit{India}$ $\rangle\}$ with $\mathit{Pr}(\mathcal{I}'_1|\mathcal{C,G})=0.12$ and $\mathcal{I}'_2=\{\langle$ $\mathit{Leander\_Paes},$ $\mathit{medals\_won},$ $\mathit{India}$ $\rangle\}$ with $\mathit{Pr}(\mathcal{I}'_2|\mathcal{C,G})=0.25$.
To find a valid instance, we need to validate $\mathcal{I}'_1 \cup \mathcal{G}$ and $\mathcal{I}'_2 \cup \mathcal{G}$ using graph constraints $\Sigma$.
We then add the instance $\mathcal{I}$ with higher joint probability into $\mathcal{G}$ when $\mathcal{I} \cup \mathcal{G} \vDash \Sigma$.
When $\forall \mathcal{I}_x, \mathcal{I}_x \cup \mathcal{G} \not\vDash \Sigma$, we will not add the instance generated by $\mathcal{C}$.
\end{example}

\section{Approximated Dynamic Repair}
\label{sec:Approx}
In this section, we analyze how to obtain a fully dynamic approach for repairing and enhancing the knowledge graph over scalable unstructured data.
We firstly discuss the dynamic common structure in Algorithm \ref{algorithm:dynamicStructure} to find and reduce the duplicated computations.
Based on the analysis of the dynamic structure, we introduce the discovery and validation through implicit graph constraints in Section \ref{sec:dynamicCommonStructure}.
The evaluation of graph similarity is introduced for the instance validation (through implicit graph constraints) (Section \ref{sec:psRepairValidation}) and linkage prediction (Section \ref{sec:link_prediction}).
To achieve a dynamic computation of approximated graph similarity, we introduce a scalable linkage-based graph embedding in Section \ref{sec:graphfraction}.
In Section \ref{sec:approxEvalGE}, we conclude the evaluations in this section and prove that the knowledge graph enhancement process is dynamic.
We also notice the cold start problems (Section \ref{sec:coldstart}) during the dynamic enhancing process and provide possible solutions.

\subsection{Dynamic Common Structure}
\label{sec:dynamicCommonStructure}

In this section, we aim to find an optimizable common structure among the three operations for dynamic repair processing i.e., instance validation, instance repair and constraints discovery.
As instance repair is converted into instance validation with the optimize-validate model in Section \ref{sec:cvRelationRepair}, the other two operations, validation and discovery, need to be analyzed and made dynamic.
We summarize constraints discovery and constraints based validation first.

\ul{\textbf{Instance Vaildation}.}
Given a set of graph constraints $\Sigma$ and graph database $\mathcal{G}$, instance validation is an error detection process to generate a set $\mathit{vio}(\mathcal{G}, \Sigma)$ with all subgraphs in $\mathcal{G}$ violating to $\Sigma$ i.e., $\mathit{vio}(\mathcal{G}, \Sigma) = \{G \subset \mathcal{G}|\exists h:G \leftrightarrow \varphi.Q, \varphi=Q[\overline{x}](X \Rightarrow Y) \in \Sigma \wedge G \not\vDash X \to Y\}$.
Given new instance $\mathcal{I}$, when $\mathit{vio}(\mathcal{G} \cup \mathcal{I}, \Sigma) = \emptyset$, there is no violation in $\mathcal{I} \cup \mathcal{G}$ given $\Sigma$ ($\mathcal{I} \cup \mathcal{G} \vDash \Sigma$), whereas no match between the subgraphs containing tuples from $\mathcal{I}$ and $\varphi.Q$ will also lead to $\mathit{vio}(\mathcal{I} \cup \mathcal{G}, \Sigma) = \emptyset$.
Since the validation $\mathcal{G}\cup\mathcal{I} \vDash \Sigma$ only checks the tuples involving matches and the constraints $\Sigma$ should update along the enhancing process, merging instance $\mathcal{I}$ according to $\mathcal{G} \cup \mathcal{I} \vDash \Sigma$ would introduce potential noisy tuples into $\mathcal{G}$.
To distinguish different validation status of tuples and instances, we extend the validation problem to the concept of (in)valid instance, which asks $\mathcal{G} \cup \mathcal{I} \vDash \Sigma$ and $\forall$ tuple $s \in \mathcal{I}, \exists G_x(s) \subset \mathcal{G} \cup \mathcal{I}$ s.t. $\exists \gamma: G_x(s) \leftrightarrow \varphi.Q, \exists \varphi \in \Sigma \wedge G_x(s) \vDash X \to Y$.


\ul{\textbf{Graph Constraint Discovery}.}
New constraints are discovered based on the support sets.
The graph constraints can be classified into two classes, positive and negative constraints.
Given positive constraint $\varphi^p=Q[\overline{x}](X \Rightarrow Y)$ and graph database $\mathcal{G}$, the support set is defined with $\mathit{Supp}(\mathcal{G}, \varphi^p) = \{G_i|\exists \gamma: G_i\leftrightarrow Q[\overline{x}], G_i \vDash X\to Y, G_i \subset \mathcal{G}\}$, where mapping $\gamma$ is a matching since subgraph pattern $Q[\overline{x}]$ should be both detectable and applicable under graph $\mathcal{G}$.
When $|\mathit{Supp}(\mathcal{G}, \varphi^p)| \geq \sigma$ and $\varphi^p$ is not trivial, $\Sigma$ will include a positive constraint $\varphi^p$.
Negative constraint $\varphi^n=Q[\overline{x}](X \Rightarrow \mathit{False})$ \cite{DBLP:conf/amw/Corte-CalabuigP12} catches the illegal patterns in a graph database.
A practical approach for finding negative constraint is to count the maximum support of a positive constraint as an extension from a negative constraint with modifications \cite{DBLP:conf/sigmod/FanHLL18}.


To conclude, the discovery of new constraints relies on the number of supports from the graph database.
The validation process of a graph database finds all matches of each constraint and checks whether the matches satisfy the corresponding logistic expression.
Therefore, graph matching is a common structure for constraint discovery and graph instance validation. 

Knowledge enhancement is a process of generating new entities and filling the topology for a knowledge graph.
While inserting new tuples into the knowledge graph, the full set of attribute information of entities can hardly be observed e.g., most of the attribute information does not exist for a newly inserted entity.
Graph constraints involving graph topology show more applicability under this case, rather than relying on entity labels.
Moreover, considering the topology constraints instead of general graph constraints would be a practical pruning for dynamic enhancement.
Two special cases of graph constraints are utilized in the rest of this work i.e., $\varphi^{p/n}=Q[\overline{x}](\emptyset \Rightarrow \mathit{True/False})$ by asserting legal and illegal graph topology patterns.
To find frequent subgraph patterns, the support set for pattern $G_x$ is defined with $\mathit{Supp}(\mathcal{G}, G_x)=\{G_i \subset \mathcal{G}| \exists \gamma: G_i \leftrightarrow G_x\}$.
For this case, a constraint $\varphi$ is trivial when $|\varphi.Q| \leq 2$, and we only consider the equivalence check among edge labels while finding a match between two graph patterns.
Lemma \ref{prop:simplifyVD_tuple} analyzes the equivalence between subgraph matching and instance validation based on graph constraints.


\begin{lemma}
\label{prop:simplifyVD_tuple}
Assuming that $\Sigma$ is the collection of graph constraints with $\forall \varphi \in \Sigma, \varphi=Q[\overline{x}](\emptyset \Rightarrow \mathit{True/False})$ for graph database $\mathcal{G} \vDash \Sigma$, we have the following conclusions:

(C1) For valid tuple $s$, at least one non-trivial subgraph pattern containing tuple $s$ has a matching degree equal or greater than $\sigma$ i.e., $\exists G_x(s) \subset \mathcal{G} \cup \{s\}, \exists \gamma: G_x(s) \leftrightarrow \varphi^p.Q, \exists \varphi^p \in \Sigma, G_x(s) \vDash \varphi^p \Leftrightarrow \exists G_x(s) \subset \mathcal{G} \cup \{s\}, |\mathit{Supp}(\mathcal{G}, G_x(s))| \geq \sigma$.

(C2) For invalid tuple $s$, at least one no support subgraph $G(s)$ leads to a support of equal or greater than $\sigma$ on non-trivial pattern $G(s)\setminus\{s\}$ i.e., $\exists G_x(s) \subset \mathcal{G} \cup \{s\}, \exists \gamma: G_x(s) \leftrightarrow \varphi^n.Q, \exists \varphi^n \in \Sigma \Leftrightarrow \exists G_x(s) \subset \mathcal{G} \cup \{s\}, |\mathit{Supp}(\mathcal{G}, G_x(s) \setminus \{s\})| \geq \sigma \wedge \forall G_x(s), \mathit{Supp}(\mathcal{G}, G_x(s))=\emptyset$. 

(C3) Tuple $s$ is unknown for current graph constraints $\Sigma$ when $s$ is not a valid or invalid tuple.

(C4) $\{s\} \cup \mathcal{G} \vDash \Sigma \Leftrightarrow s$ is unknown or valid. 
\end{lemma}

\begin{proof}
According to the discovery problem of graph constraints, we have $\forall \varphi^p \in \Sigma$, $|\mathit{Supp}(\mathcal{G}, \varphi^p)| = |\mathit{Supp}(\mathcal{G}, Q[\overline{x}])| \geq \sigma$.
Also, for each negative graph constraint $\varphi^n$, we have $\forall \varphi^n \in \Sigma$, $\max_{\varphi^{n}_i \in \Phi^n} |\mathit{Supp}(\mathcal{G}, \varphi^{n}_i)|=\max_{\varphi^n.Q'}|\mathit{Supp}(\mathcal{G}, \varphi^n.Q')| \geq \sigma$, where $\Phi^n$ is the variant space of $\varphi^n$ and $\varphi^n.Q'$ is a subgraph with minimum change to $\varphi^n.Q$.
According to the description of incremental instance validation, when $\mathcal{G} \vDash \Sigma$, we have $\mathcal{G} \cup \{s\} \vDash \Sigma$ $\Leftrightarrow$ $\forall G_x(s) \subset \mathcal{G} \cup \{s\}$, $\not\exists \gamma: G_x(s) \leftrightarrow \varphi^n.Q, \varphi^n \in \Sigma$.

Since the support set $\mathit{Supp}(\mathcal{G}, G_x)$ is calculated based on isomorphism $\gamma_\mathit{x,i}: G_x \leftrightarrow G_i \subset \mathcal{G}$, 
we have $\exists \gamma_\mathit{x, i}: G_x(s) \leftrightarrow \varphi^p_i.Q, \varphi^p_i \in \Sigma$ $\Rightarrow$ $|\mathit{Supp}(\mathcal{G}, G_x(s))| = |\mathit{Supp}(\mathcal{G}, \varphi^p_i)| \geq \sigma$. (This is because $\forall G \in \mathit{Supp}(\mathcal{G}, \varphi^p_i), \exists \gamma: \varphi^p_i.Q \leftrightarrow G$ leads to both $\gamma_\mathit{x, i} \cdot \gamma: G_x(s) \leftrightarrow G$ and $G \in \mathit{Supp}(\mathcal{G}, G_x(s))$).
Also, we have $\exists G_x(s), |\mathit{Supp}(\mathcal{G}, G_x(s))| \geq \sigma$ $\Rightarrow$ $\exists \varphi^p \in \Sigma, \exists \gamma: G_x(s) \leftrightarrow \varphi^p.Q$.
(This is because $\forall G_i \in \mathit{Supp}(\mathcal{G}, G_x(s)), \exists \gamma_i: G_x(s) \leftrightarrow G_i$ means $\gamma_\mathit{ij}=\gamma_i^{-1} \cdot \gamma_j: G_i \leftrightarrow G_j$ and $|\mathit{Supp}(\mathcal{G}, G_i)| \geq \sigma$. 
Thus, we have a positive graph constraint $\varphi^p \in \Sigma$ whose graph pattern $\varphi^p.Q$ is isomorphic to $G_i$ by isomorphism $\gamma_\mathit{p,i}$. 
Since $\gamma_i$ is an isomorphism between $G_x(s)$ and $G_i$, we have $\gamma=\gamma_i \cdot \gamma_\mathit{p,i}^{-1}$ as the isomorphism between $G_x(s)$ and $\varphi^p.Q$).
Thus, we have C1 about a valid tuple $s$.

When tuple $s$ is a detected violation, we have $\exists G_x(s) \subset \mathcal{G} \cup \{s\}$ such that $\exists \gamma_\mathit{x, i}: G_x(s) \leftrightarrow \varphi^n_i.Q, \varphi^n_i \in \Sigma$.
Assume that $\mathcal{G} \vDash \Sigma$, we have $\forall s_i \in \mathcal{G}$, (1) $|\mathit{Supp}(\mathcal{G}, G(s_i))| < \sigma$ $\wedge |\mathit{Supp}(\mathcal{G}, G(s_i)\setminus\{s_i\})|<\sigma$ or (2) $|\mathit{Supp}(\mathcal{G}, G(s_i))| \geq \sigma$.
When $G(s)$ is a match of $\varphi^n$, removing tuples other than $s$ will not lead to a frequent pattern, therefore, we have $|\mathit{Supp}(\mathcal{G}, G(s)\setminus\{s\})| \geq \sigma$ as shown in C2.
\end{proof}

We extend Lemma \ref{prop:simplifyVD_tuple} to the instance validation on $\mathcal{I}$.

\begin{proposition}
\label{prop:simplifyVD}
Given $\Sigma$ as the collection of positive and negative graph constraints, we have the following conclusions:

(C1) $\mathcal{G} \cup \mathcal{I} \vDash \Sigma$ $\Leftrightarrow$ $\forall s \in \mathcal{I}, s$ is unknown or valid.

(C2) $\mathcal{I}$ is a valid instance $\Leftrightarrow$ $\forall s \in \mathcal{I}, s$ is a valid tuple.

\end{proposition}

Based on Proposition \ref{prop:simplifyVD}, we can validate a tuple or an instance through the number of supporting subgraphs in graph database instead of explicitly discovering the constraints.
Observing this, the dynamic problems in graph repair are solvable through handling the instance validation problem based on implicit constraints.

\subsection{Implicit Constraint for Instance Validation}
\label{sec:psRepairValidation}

We introduce this section to achieve dynamic instance validation through implicit graph constraints.
As Lemma \ref{prop:simplifyVD_tuple} shows, when $\exists G(s), |\mathit{Supp}(\mathcal{G}, G(s))|\geq \sigma$, validating tuple $s \in \mathcal{I}$ is equal to validating it with graph constraints.

To find candidates, enumerating a subgraph $G(s)$ containing tuple $s$ consumes $O(|\mathcal{G}\cup\mathcal{I}|^{q_m-2})$ when we consider a maximum subgraph size $q_m, q_m>2$.
For finding the support set of $G(s)$, we need to find its isomorphic subgraphs from $\mathcal{G}$ with $O(f(q_m)*|\mathcal{G}|^{q_m})$, where $f$ is an exponential function of $q_m$.
The instance validation with $T_{\mathit{valid}}(\mathcal{I})=O(f(q_m)*|\mathcal{G}\cup\mathcal{I}|^{2q_m-2})$ is fixed-parameter tractable.
$T_{\mathit{valid}}$ leads to a lower bound of amortized time $\tilde{O}(\frac{f(q_m)*|\mathcal{G}\cup\mathcal{I}|^{2q_m-2}}{|\mathcal{I}|})$ for Algorithm \ref{algorithm:dynamicStructure} showing the hardness of dynamic processing even when $q_m$ is a constant.
The hardness mainly comes from the enumeration of subgraphs and graph matching, where subgraph enumeration relates to two sides i.e., enumerating $G(s) \subset \mathcal{G} \cup \mathcal{I}$ containing tuple $s$ for validation and subgraphs $G_i \subset \mathcal{G}$ as candidates.

To solve dynamic hardness, we start by reducing graph pattern enumeration $G(s) \subset \mathcal{G} \cup \mathcal{I}$.
As knowledge graph contains the structural information on tuple $s$, the \rdf tuples connecting to $s$ within limited steps will have a closer relationship with $s$ \cite{DBLP:journals/corr/abs-2002-00819}.
Observing this, we restrict our subgraph searching space onto the graph patterns that contain both tuple $s$ and the tuples within a fixed number of steps instead of enumerating all the subgraphs $G(s)$.
We have Definition \ref{def:localizedpattern} for localized patterns to formalize this, where $\mathit{dist}(v_1,v_2)$ computes the minimum \textsc{Depth First Search} length between $v_1$ and $v_2$ without considering the direction of edges.
Proposition \ref{prop:localized_pattern_size_ub} considers the upper bound of the size of localized patterns.
Lemma \ref{lemma: supp_localize_pattern} prunes the searching space of finding matching subgraphs $G_i$ from $\mathcal{G}$.

\begin{definition}
\label{def:localizedpattern}
Given length $\mathit{l}$ and an \textsc{RDF} tuple $\langle h,r,t \rangle$ in graph database $\mathcal{G}_x$, a localized pattern $\mathsf{P}_{\langle h,r,t \rangle}^l=\{V_\mathsf{P},E_\mathsf{P},L_\mathsf{P}\}$ is defined by
$V_\mathsf{P} = \{v \in \mathcal{G}_x.V|\mathit{dist}(v,h)\leq l \wedge \mathit{dist}(v,t)\leq l\}$
$E_\mathsf{P} = \{e(v_1,v_2)\in \mathcal{G}_x.E| v_1,v_2 \in V_\mathsf{P}\}$, $L_\mathsf{P}=\mathcal{G}_x.L$.
\end{definition}

\begin{example}
\label{exp:patterndef}
Consider Example \ref{exp:intro} and Figure \ref{fig:demo}, we give an example about localized pattern here.
When taking length $l=1$ and tuple $s=\langle \mathit{India},\mathit{contains},\mathit{Gorakhpur} \rangle$, the corresponding localized pattern is $\mathsf{P}^1_{s} = \{V_{\mathsf{P}}, E_{\mathsf{P}}, L_{\mathsf{P}}\}$, with $V_{\mathsf{P}} = $ $\{\mathit{India},$ $\mathit{Gorakpur},$ $\mathit{Earth},$ $\mathit{Gurgaon},$ $\mathit{Sikkim},$ $\mathit{ZTE},$ $\mathit{Leander\_Paes},$ $\mathit{Uttar\_Pradesh},$ $\mathit{Anurag\_Kashyap}\}$.
The set $E_{\mathsf{P}}$ is the collection of directed edges between entities in $V_{\mathsf{P}}$.
\end{example}

\begin{proposition}
\label{prop:localized_pattern_size_ub}
Given maximum out-degree $d_{M+}$ and in-degree $d_{M-}$ for graph $\mathcal{G}$, $|\mathsf{P}^l| \leq 2(d_{M+}+d_{M-})^l+2=2d_M^l+2$. 
\end{proposition}

\begin{lemma}
\label{lemma: supp_localize_pattern} 
Given tuple $s=\langle h,r,t\rangle \in \mathcal{I}$ and 
the support set 
$\overline{\mathit{Supp}}(\mathcal{G}, \mathsf{P}^l_s) = \{G_i \subset \mathcal{G}|\exists \gamma: G(s) \leftrightarrow G_i, G(s) \subseteq \mathsf{P}^l_s\}$ for localized pattern $\mathsf{P}^l_s$ , we have 
$\overline{\mathit{Supp}}(\mathcal{G}, \mathsf{P}^l_s) \equiv \{G_i \subseteq \mathsf{P}^l_{s'(r)}|\exists \gamma: G_i \leftrightarrow G(s) \subseteq \mathsf{P}^l_s, s'(r) \in \mathcal{G}\}$.
\end{lemma}

\begin{proof}
For each $G_i \in \overline{\mathit{Supp}}(\mathcal{G}, \mathsf{P}^l_s)$, we have $\exists \gamma: G(s) \leftrightarrow G_i, G(s) \subseteq \mathsf{P}^l_s$.
Since $\forall v \in G(s) \subseteq \mathsf{P}^l_s, \mathit{dist}(v, h) \leq l \wedge \mathit{dist}(v, t) \leq l$, for each undirected path linking $v$ and $h, t$ written as $\mathit{Path}(v, h)=v,v_1...h$ and $\mathit{Path}(v, t)=v,v'_1,...t$, we have $\gamma(\mathit{Path}(v, h)) = \gamma(v),\gamma(v_1),...\gamma(h)$ as a path in $G_i$, which indicates $|\mathit{Path}(\gamma(v), \gamma(h))| \leq l$ and $|\mathit{Path}(\gamma(v), \gamma(t))| \leq l$.
Thus, $G_i \subseteq \mathsf{P}^l_{s'(r)}, s'(r) = \langle \gamma(h), r, \gamma(t) \rangle \in \mathcal{G}$.
\end{proof}

Lemma \ref{lemma: supp_localize_pattern} suggests that we could index all occurrence of $r$ in $\mathcal{G}$ and find the subgraphs as support.
We use Lemma \ref{lemma: anti-monotonic_supp} to show the anti-monotonic properties observed in the support set of localized patterns.

\begin{lemma}
\label{lemma: anti-monotonic_supp}
Given $\mathsf{P}^l_s \subset \mathcal{G}\cup\mathcal{I}$ as the localized pattern for tuple $s \in \mathcal{I}$, we have (C1) $\forall G_x \in \overline{\mathit{Supp}}(\mathcal{G}, \mathsf{P}^{l+1}_s), \exists G_{x, i} \subseteq G_x$ s.t. $\exists \gamma_{i, j}: G_{x, i} \leftrightarrow G_j, \exists G_j \in \overline{\mathit{Supp}}(\mathcal{G}, \mathsf{P}^l_s)$ and (C2) $|\overline{\mathit{Supp}}(\mathcal{G}, \mathsf{P}^l_s)| \leq |\overline{\mathit{Supp}}(\mathcal{G}, \mathsf{P}^{l+1}_s)|, \forall l \geq 1$.
\end{lemma}

\begin{proof}
C1 concludes that a support subgraph for $\mathsf{P}^{l+1}_s$ contains a matching pattern from $\mathsf{P}^l_s$.
According to Lemma \ref{lemma: supp_localize_pattern}, $\forall G_x \in \overline{\mathit{Supp}}(\mathcal{G}, \mathsf{P}^{l+1}_s), \exists \gamma_{x, y}: G_x \leftrightarrow G_y, G_y \subseteq \mathsf{P}^{l+1}_{s'}$ $\wedge \forall v \in G_y.V, |\mathit{Path}(v, s'.h)| \leq l+1, |\mathit{Path}(v, s'.t)| \leq l+1$.
We discuss different situations based on the paths in $G_x$ (1) when $\max_{(v_i, v_j) \in  G_x.V \times \{s'.h, s'.t\}} |\mathit{Path}(v_i, v_j)| \leq l$, $G_x$ is a support subgraph and $G_x \in \overline{\mathit{Supp}}(\mathcal{G}, \mathsf{P}^{l}_s)$; (2) when $\max_{(v_i, v_j) \in  G_x.V \times \{s'.h, s'.t\}} |\mathit{Path}(v_i, v_j)| = l+1$, for each path with $v, v_1, ...v_{l-1}, (s'.h|s'.t)$, removing $v$ will lead to a subgraph $G'_x \subsetneq G_x \subseteq \mathsf{P}^{l+1}_{s'} \wedge G'_x \subseteq \mathsf{P}^l_{s'}$, which indicates a graph matching between $G'_x$ and $\exists G_y \subseteq \mathsf{P}^l_s$.
Since $\overline{\mathit{Supp}}(\mathcal{G}, \mathsf{P}^l_s) \subseteq \overline{\mathit{Supp}}(\mathcal{G}, \mathsf{P}^{l+1}_s)$, we have C2 for numerical connections among various $l$.
\end{proof}

To study the instance validation problem based on localized patterns, we introduce Proposition \ref{prop:localized_pattern_supp_legal} and \ref{prop:localized_pattern_supp_invalid} to draw its connections to $\mathcal{G} \cup \mathcal{I} \vDash \Sigma$ (legal instance) and invalid instance, respectively. 

\begin{proposition}
\label{prop:localized_pattern_supp_legal}
$\forall s \in \mathcal{I}, \overline{\mathit{Supp}}(\mathcal{G}, \mathsf{P}^l_s) \neq \emptyset \Rightarrow \mathcal{G}\cup\mathcal{I} \vDash \Sigma$.
\end{proposition}

\begin{proof}
Since $\emptyset \neq \overline{\mathit{Supp}}(\mathcal{G}, \mathsf{P}^l_s) \subseteq \bigcup_{G(s) \subset \mathcal{G} \cup \mathcal{I}} \mathit{Supp}(\mathcal{G}, G(s))$, the enhanced graph database $\mathcal{G} \cup \mathcal{I}$ is not invalid.
Thus, we have $\forall s \in \mathcal{I}, \overline{\mathit{Supp}}(\mathcal{G}, \mathsf{P}^l_s) \neq \emptyset \Rightarrow \mathcal{G}\cup\mathcal{I} \vDash \Sigma$.
\end{proof}

\begin{proposition}
\label{prop:localized_pattern_supp_invalid}
When $\sigma=1$, $\overline{\mathit{Supp}}(\mathcal{G}, \mathsf{P}^l_s)=\emptyset \wedge \exists G_i(s) \subseteq \mathsf{P}^l_s, |G_i(s)| > 3$ s.t. $\mathit{Supp}(\mathcal{G}, G_i(s) \setminus \{s\})\neq \emptyset$ $\Leftrightarrow s$ is invalid.
\end{proposition}

\begin{proof}
We first prove the assertion $\overline{\mathit{Supp}}(\mathcal{G}, \mathsf{P}^l_s)=\emptyset \Leftrightarrow \forall G(s)\subset\mathcal{G}\cup\mathcal{I}, \mathit{Supp}(\mathcal{G}, G(s)) = \emptyset$. 
Since $\forall G(s) \in \mathcal{G} \cup \mathcal{I}, \mathit{Supp}(\mathcal{G}, G(s)) = \emptyset \Leftrightarrow \bigcup_{G(s) \subset \mathcal{G} \cup \mathcal{I}} \mathit{Supp}(\mathcal{G}, G(s)) = \overline{\mathit{Supp}}(\mathcal{G}, \mathsf{P}^{l \to |\mathcal{G} \cup \mathcal{I}|< \infty}_s) = \emptyset$, we have $\overline{\mathit{Supp}}(\mathcal{G}, \mathsf{P}^{l}_s) = \emptyset, l<|\mathcal{G} \cup \mathcal{I}|$ based on Lemma \ref{lemma: anti-monotonic_supp}. When $\exists G_i(s) \subseteq \mathsf{P}^l_s, |G_i(s)| > 3$, $G_i(s) \setminus \{s\}$ is a non-trivial pattern. 
Thus, $\exists G(s)=G_i(s), |\mathit{Supp}(\mathcal{G}, G(s) \setminus \{s\})| \geq \sigma=1 \wedge \forall G(s) \subset \mathcal{G} \cup \mathcal{I}, \mathit{Supp}(\mathcal{G}, G(s)) = \emptyset \Leftrightarrow \exists s \in \mathcal{I}, s$ is a invalid tuple based on Lemma \ref{prop:simplifyVD_tuple}(C2).
\end{proof}

To validate instance $\mathcal{I}$, we calculate the support set $\overline{\mathit{Supp}}(\mathcal{G}, \mathsf{P}^l_s) = \{G_i \subseteq \mathsf{P}^l_{s'(s.r)}|\exists h: G_i \leftrightarrow G(s) \subseteq \mathsf{P}^l_s, \mathsf{P}^l_{s'(s.r)} \in S(\mathcal{G}, s.r)\}$
, where $\mathsf{P}^l_s \subset \mathcal{G} \cup\mathcal{I}$ and $S(\mathcal{G}, r) = \{\mathsf{P}^l_{s'(r)} \subset \mathcal{G}| s'(r) \in \mathcal{G}\}$ is the localized pattern set with all occurrences of relation $r$ in $\mathcal{G}$.
The amortized time complexity for validation using implicit constraints is $\tilde{O}(\frac{g(l)\sum_{s\in\mathcal{I}}|S(\mathcal{G}, s.r)|}{|\mathcal{I}|})= \tilde{O}(g(l)\cdot\max_{r\in L_\mathcal{E}} |S(\mathcal{G}, r)|)$, where $g(l)=2^{d_M^l+d_M^{2l}}$ is the complexity of finding matching subgraphs  between two localized patterns.
Considering both $l$ and $\max_{r\in L_\mathcal{E}} |S(\mathcal{G}, r)|$ as constants, the validation problem is fixed-parameter tractable and has an amortized time complexity of $\tilde{O}(1)$.

However, the amortized complexity indicates that such a validation process is not applicable to general cases such as dense graphs.
Firstly, for $O(g(l))=O(2^{d_M^{2l}})$, enumerating subgraphs and matches are both time-consuming when the degree of each vertex is high.
We consider an approximated localized pattern matching in Section \ref{sec:graphfraction} with an amortized hardness $\tilde{O}(d_M^l)$.
Secondly, $\max_{r \in L_\mathcal{E}} |S(\mathcal{G}, r)|$ is a large constant and rises with the size of graph database $\mathcal{G}$.
To avoid computation over large candidate sets, we pick a sampling subset of $S(\mathcal{G}, r)$ denoted as $S^*(\mathcal{G}, r)$.
In this way, instance validation is a dynamic process without high amortized time consumption.

\subsection{Linkage Prediction}
\label{sec:link_prediction}

We introduce this section to generate linkage prediction $\mathit{Pr}(\mathcal{I}|\mathcal{G})$ for $\mathcal{\mathcal{M}_{\mathit{lp}}}$ in Section \ref{sec:cvRelationRepair}.

\ul{\textbf{Linkage Prediction}.}
Given a tuple $\langle h,t \rangle$ in graph database $\mathcal{G}$, for each relation label $r\in L_\mathcal{E}$, linkage prediction generates the probability $\mathit{Pr}(r|\langle h,t \rangle, \mathcal{G})$ based on the graph database.

Since the localized pattern $\mathsf{P}^l_{s(r)}$ collects the correlated relation labels with $r$ in a graph database, the graph similarity between two localized patterns ($\mathsf{P}^l_{s(r)}$ and $\mathsf{P}^l_{s'(r)}$) reveals the consistency of the context associated with relation label $r$.
Thus, given entities $\langle h, t \rangle$ and a possible relation $r$ between $h$ and $t$, the similarity between $\mathsf{P}^l_{\langle h, r, t \rangle}$ and $\mathsf{P}^l_{s'(r)} \in S(\mathcal{G}, r)$ shows the consistency support from ground truth $s' \in \mathcal{G}$, which can be taken as an estimation of linkage prediction $\mathit{Pr}(r|\langle h,t \rangle, \mathcal{G})$.
The classic metrics for similarity evaluation such as \textsc{Jaccard} and \textsc{Cosine} compare the intersection and union of two patterns.
When $|\mathsf{P}^l_{s'}| \ll |\mathsf{P}^l_s|$, given subgraphs $G_i(s') \subseteq \mathsf{P}^l_{s'}, G_j(s) \subseteq \mathsf{P}^l_{s}$ with $\exists \gamma: G_i(s') \leftrightarrow G_j(s)$, we have $|G_i(s')| = |G_j(s)|$, which represents an intersection between $\mathsf{P}^l_{s}$ and $\mathsf{P}^l_{s'}$ i.e., $|G_i(s')|=|G_j(s)| \leq \min(|\mathsf{P}^l_{s'}|, |\mathsf{P}^l_{s}|) \ll |\mathsf{P}^l_s|$, which means that the classic similarity measurements can not capture the patterns with different sizes effectively.
To capture the consistency of two localized patterns, we introduce Definition \ref{def:sim_min} as a similarity mertic.

\begin{definition}
\label{def:sim_min}
Given two localized patterns $\mathsf{P}^l_{s_1}, \mathsf{P}^l_{s_2}$ with the same centric relation label $s_1.r = s_2.r$,
the similarity $\mathit{Sim}(\mathsf{P}^l_{s_1}, \mathsf{P}^l_{s_2})$ is equal to
  
$$\mathit{Sim}(\mathsf{P}^l_{s_1}, \mathsf{P}^l_{s_2}) = \frac{\max_{h: G_i(s_1) \leftrightarrow G_j(s_2)} |h| }{\min(|\mathsf{P}^l_{s_1}|, |\mathsf{P}^l_{s_2}|)}, $$
  
where $|h|$ is the number of matching pairs between $G_i(s_1) \subseteq \mathsf{P}^l_{s_1}$ and $G_j(s_2) \subseteq \mathsf{P}^l_{s_2}$.
\end{definition}


Linkage prediction has been analyzed in many works \cite{DBLP:conf/cikm/Liben-NowellK03,DBLP:conf/nips/Kazemi018,DBLP:conf/kdd/LeroyCB10}.
They share the same idea of comparing the similar sub-structure in graph database. 
In this work, we find a portable linkage prediction method to optimize the relation extraction results dynamically.
Based on Definition \ref{def:sim_min}, we have 

\begin{align*}
\mathit{Pr}(r|\langle h,t \rangle, \mathcal{G}) &= \mathbb{E}_{\langle h,r,t \rangle|\mathcal{G}} \mathit{Sim}(\mathsf{P}^l_{\langle h,r,t \rangle}, \mathsf{P}^l_{s'(r)}) \\
&\approx \frac{1}{|S^*(\mathcal{G}, r)|} \sum_{\mathsf{P}^l_{s'} \in S^*(\mathcal{G}, r)}\mathit{Sim}(\mathsf{P}^l_{\langle h, r, t \rangle}, \mathsf{P}^l_{s'}).
\end{align*}

Based on Section \ref{sec:cvRelationRepair}, we could optimize and validate the relation labels with \textsc{Top}-k knowledge acquisition resuls w.r.t. a sentence $c$ and a pair of entities $\langle h, t \rangle$.
Given parameter $k$ as \textsc{Top}-k relation labels from $\mathcal{M}_{\mathit{ka}}$ and graph database $\mathcal{G}$, we have the amortized time complexity $\tilde{O}(k\cdot g(l)\max_{r\in L_{\mathcal{E}}}|S^*(\mathcal{G}, r)|)$ leading to a dynamic approach.

\subsection{Approximated Graph Matching}
\label{sec:graphfraction}

To tackle the hardness brought by finding matching subgraphs from localized patterns, we introduce this section for approximated graph matching, which enables the instance validation and linkage prediction over scalable \rdf data.

Given localized pattern $\mathsf{P}^l_s$, we aim to find a set of matched localized patterns for instance validation and linkage prediction. 
According to Proposition \ref{prop:localized_pattern_supp_legal} and \ref{prop:localized_pattern_supp_invalid}, a matched pattern $\mathsf{P}^l_{s'}$ for $\mathsf{P}^l_s$ need to satisfy the following conditions: (1) $s.r = s'.r$ and $\exists \gamma: G_i(s) \leftrightarrow G_j(s'), G_i(s) \subseteq \mathsf{P}^l_s, G_j(s') \subseteq \mathsf{P}^l_{s'}$ (i.e., $\gamma(s) = s' \in \mathsf{P}^l_{s'}$); (2) the isomorphism subgraph $G_j(s)$ is not trivial (i.e., $|G_j(s')| \geq 3$). 
Moreover, considering Proposition \ref{prop:localized_pattern_supp_invalid} and Definition \ref{def:sim_min}, we have certain restrictions on the matching algorithm: (1) being scalable to dynamic graphs; (2) being able to compare the consistency of two localized patterns of different sizes.
Existing approaches find matched subgraphs through graph representation learning \cite{DBLP:journals/corr/abs-1909-00958,DBLP:journals/corr/abs-2002-00388}, whereas dynamic graph representation surveyed in \cite{DBLP:journals/jmlr/KazemiGJKSFP20} shows no applicability to the restrictions discussed above.

Graph embedding algorithm TraverseR(TR) traverses the relation labels in a localized pattern with fixed length and records the paths into a set.
To find the graph embedding on localized pattern topology and focus on central tuple, TraverseR uses Algorithm \ref{algorithm:graphEmbedding} to generate an embedding $\mathbb{M}$ that maps all relation paths containing central relation $r$ of length $l$ onto their occurrences as weight for $\mathsf{P}^l_{\langle h,r,t \rangle}$.

\begin{algorithm}[h]
	\small
	\caption{TraverseR(TR): central relation focused embedding} %
	\label{algorithm:graphEmbedding} %
	\begin{algorithmic}[1]
		\REQUIRE Graph pattern $\mathsf{P}^l(h,r,t)=(V_p,E_p,L_p)$, Length $\mathit{l}$.
		\ENSURE Graph embedding $\mathbb{M}$.
		\STATE $\mathsf{S}_\mathit{head}:=\{\}, \mathsf{S}^+_\mathit{head}:=\{([r],h)\}$
		\FOR{maximum pattern length $\mathit{l}_m:=1$ count to $\mathit{l}$}
		\STATE $\mathsf{S}_\mathit{temp} := \{\}$
		\FOR{each $(\mathit{Path}, v) \in \mathsf{S}^+_\mathit{head}$}
		\FOR{each $e(v,v') \in E_p-\{e(h,t)\}$}
		\STATE $\mathsf{S}_\mathit{temp} = \mathsf{S}_\mathit{temp}\cup (\mathit{Path} \cup L_p(e(v,v')), v')$
		\ENDFOR
		\ENDFOR
		\STATE $\mathsf{S}_\mathit{head} = \mathsf{S}_\mathit{head}\cup\mathsf{S}^+_\mathit{head}$
		\STATE $\mathsf{S}^+_\mathit{head} = \mathsf{S}_\mathit{temp}$
		\ENDFOR
		\STATE $\mathsf{S}_\mathit{head} = \mathsf{S}_\mathit{head}\cup\mathsf{S}^+_\mathit{head}$
		\STATE $\mathsf{S}_\mathit{tail}:=\{([r],t)\}, \mathsf{S}_\mathit{tail}^+:=\{([r],t)\}$
		\STATE run tail side with Line 2-9.
		\STATE Operate \textsc{Map-Reduce} sentences: Line 13-17.
		\STATE $\mathsf{S}_\mathit{tail}.\mathit{map}((\mathit{Pat}, \mathit{SP}) \to (\mathit{Pat.size}, \mathit{Pat}))$.
		\STATE $\mathbb{M} = \mathsf{S}_\mathit{head}.\mathit{map}((\mathit{Pat},\mathit{SP}) \to (l-\mathit{Pat.size}+1, \mathit{Pat}))$
		\STATE $\mathbb{M} = \mathbb{M}.\mathit{join}(\mathsf{S}_{tail})$
		\STATE $\mathbb{M} = \mathbb{M}.\mathit{map}((\mathit{TSize},(\mathit{PatH},\mathit{PatT})) \to (\mathit{PatH} \cup \mathit{PatT}, 1))$
		\STATE $\mathbb{M} = \mathbb{M}.\mathit{reduceByKey}(\_+\_).\mathit{collect}()$
		\RETURN $\mathbb{M}$
	\end{algorithmic}
\end{algorithm}

Breadth first search is applied to traverse all possible paths.
TraverseR constructs graph embedding starting from the head and tail entities with an initialized extracting queue $\mathsf{S}$ ($\mathsf{S}_\mathit{head}$ or $\mathsf{S}_\mathit{tail}$) in Line 1.
We introduce variable $\mathsf{S}^+$ and $\mathsf{S}_{\mathit{temp}}$ to hold the next step candidate and current temporary value, respectively.
The paths are traversed from head and tail entity separately.
Queues $\mathsf{S},\mathsf{S}^+,\mathsf{S}_{\mathit{temp}}$ maintains tuples in the form of $(\mathit{Path}, \mathit{StartPoint})$, which denotes the paths we already obtained and the next start point to make a further step
e.g., when starting from head entity $h$ and tail $t$, the initial value of $\mathsf{S}^+$, $\mathsf{S}^+_{head}$ is $([r], h)$ and $\mathsf{S}^+_\mathit{tail}$ is $([r],t)$ (Line 1, 10).
$l_m$ denotes the maximum length of paths in queue $\mathsf{S}$ and ranges in $[1, l]$.
When making a further step, we add the relation label visited into paths and update the next start point (Line 6).
We keep all historic value in $\mathsf{S}$ with $\mathsf{S} := \mathsf{S}\cup\mathsf{S}^+$ (Line 7, 9).
After one iteration, we update $\mathsf{S}^+$ with current temporary value as candidate for next searching process (Line 8).
In this way, all the paths from $h$ or $t$ containing $r$ is extracted with different lengths.

To combine both sides of extracting queue to form a complete embedding, TraverseR applies a \textsc{Map-Reduce} process joining the results provided by both sides (Line 13-17).
The \textsc{Map-Reduce} sentences are described with \textsc{Scala} grammar.
Among Line 13-17, we use $\mathit{Pat}$, $\mathit{SP}$ and $\mathit{Size}$ to denote path, start point and path size, respectively.
Plus, we use $H$ and $T$ for paths starting from head and tail.
Line 13 constructs $\mathsf{S}_\mathit{tail}$ into (Path Size, Path), and we no longer need start point during combining process.
To aid the join operation and find paths with fixed length, Line 14 initiates answer $\mathbb{M}$ with $\mathsf{S}_\mathit{head}$, which maps the paths from head and their size ($l+1$ - \textit{Path Size}, \textit{Path}).
Line 15-16 joins $\mathsf{S}_{head}$, $\mathsf{S}_{tail}$ and combines paths from head and tail set by merging paths from two sides.
Line 17 calculates the occurrences of paths and collects the results.

\begin{example}
	Consider Example \ref{exp:patterndef} and Figure \ref{fig:demo}.
	We could obtain the embedding of $\mathsf{P}^1_{s}$,$s=\langle \mathit{India},\mathit{contains},\mathit{Gorakhpur} \rangle$ by TraverseR.
	In $\mathit{TR}(\mathsf{P}^1_{s})$, we use the abbreviation of those relation labels: contains(C), containsBy(CB), admin children(AC), administrative parent(AP), country(CU), place of birth(PB).
	We have
	$\mathit{TR}(\mathsf{P}^1_{s})= \{$((C,CB), 1), ((C,C), 1), ((AC,C), 1), ((AP,C), 1),((C,CU), 1), ((C,PB), 1)$\}$
	as localized pattern embedding.
\end{example}

To capture patterns with different sizes, The following equation calculates the similarity of the embeddings of localized patterns and compares the consistency of two localized pattern instead of using Definition \ref{def:sim_min}.
Given localized patterns $\mathit{TR}(\mathsf{P}^l_{s_1})$ and $\mathit{TR}(\mathsf{P}^l_{s_2})$, we have the similarity 

\begin{align*}
\mathsf{sim}(\mathsf{P}^l_{s_1}, \mathsf{P}^l_{s_2}) = \frac{|\mathit{TR}(\mathsf{P}^l_{s_1}) \cap \mathit{TR}(\mathsf{P}^l_{s_2})|}{\min(|\mathit{TR}(\mathsf{P}^l_{s_1})|,|\mathit{TR}(\mathsf{P}^l_{s_2})|)}.
\end{align*}

As the embedding $\mathit{TR}(\mathsf{P}^l_{s})$ is a multi-set mapping paths and their occurrences, the intersection records the common paths of the two embeddings.
The size of a multi-set is calculated with $|\mathbb{M}|=\sum_{\mathit{path}\in\mathbb{M}} \mathbb{M}[\mathit{path}]$.
Proposition \ref{prop:not_trivial_iso} suggests that the similarity indicates the existence of an isomorphism between non-trivial subgraphs of two localized patterns.
To show dynamic, Proposition \ref{lemma:TRcomplexity} analyzes the time and space complexity of our embedding algorithm.

\begin{proposition}
\label{prop:not_trivial_iso}
Given localized patterns $\mathsf{P}^l_{s_1}$ and $\mathsf{P}^l_{s_2}$ with $s_1.r=s_2.r$, we have: $\mathsf{sim}(\mathsf{P}^l_{s_1}, \mathsf{P}^l_{s_2}) > 0$ $\Leftrightarrow$ $\exists \gamma: G_i(s_1) \leftrightarrow G_j(s_2), G_i(s_1) \subseteq \mathsf{P}^l_{s_1}, G_j(s_2) \subseteq \mathsf{P}^l_{s_2}$ with $\gamma(s_1) = s_2 \wedge |G_i(s_1)|=|G_j(s_2)| \geq 3, \forall l \geq 1$.
\end{proposition}


\begin{proposition}
\label{lemma:TRcomplexity}
Given a localized pattern $\mathsf{P}^l_t$ with maximum degree $d_M$ and parallelism degree $n~(l\ll n)$, TraverseR runs in $O(d_M^l/n)$, and the space complexity of $\mathbb{M}$ is $O(l^2 \cdot d_M^l)$.
\end{proposition}

When taking approximated matching patterns as candidate for instance validation, the support set is defined with $\overline{\mathit{Supp}}'(\mathcal{G}, \mathsf{P}^l_s) = \{\mathsf{P}^l_{s_i} \in S(\mathcal{G}, s.r)| \mathsf{sim}(\mathsf{P}^l_s, \mathsf{P}^l_{s_i})>0\}$.
To validate instance through matching patterns, Proposition \ref{prop:approx_supp_legal} and \ref{prop:approx_supp_invalid} conclude the conditions of legal instance and invalid instance respectively based on Proposition \ref{prop:not_trivial_iso}.

\begin{proposition}
\label{prop:approx_supp_legal}
$\forall s \in \mathcal{I}, \overline{\mathit{Supp}}'(\mathcal{G}, \mathsf{P}^l_s) \neq \emptyset \Rightarrow \mathcal{I}$ is legal.
\end{proposition}

\begin{proposition}
\label{prop:approx_supp_invalid}
Given $s=\langle h,r,t \rangle \in \mathcal{I}, \overline{\mathit{Supp}}'(\mathcal{G}, \mathsf{P}^l_s) = \emptyset$, $\sigma=1 \wedge l=1$ (1) when $\exists s'=\langle h,r',t \rangle \in \mathcal{G} \cup \mathcal{I}$, $\overline{\mathit{Supp}}'(\mathcal{G}, \mathsf{P}^l_{s'}) \neq \emptyset \Rightarrow s$ is invalid; (2) when $\nexists s'=\langle h,r',t \rangle \in \mathcal{G} \cup \mathcal{I}$, and (2.1) $\exists s_h=\langle h,r',t' \rangle \in \mathcal{G} \cup \mathcal{I}, \overline{\mathit{Supp}}'(\mathcal{G}, \mathsf{P}^{l}_{s_h}) \neq \emptyset \Rightarrow s$ is invalid, (2.2) $\exists s_t=\langle t,r',t' \rangle \in \mathcal{G} \cup \mathcal{I}, \overline{\mathit{Supp}}'(\mathcal{G}, \mathsf{P}^{l}_{s_t}) \neq \emptyset \Rightarrow s$ is invalid.
\end{proposition}

\subsection{Approximated Evaluations via Graph Embedding}
\label{sec:approxEvalGE}

We introduce this section to apply graph matching onto the proposed evaluations as a conclusion to the above four sections.
As discussed in Section \ref{sec:overviewDynamic}, dynamic hardness is noticed on three operations i.e., constraints based validation, constraints based repair, and the discovery of new constraints.
We summarize the evaluations according to Algorithm \ref{algorithm:dynamicStructure}.

Firstly, Line 4-6 in Algorithm \ref{algorithm:dynamicStructure} involves fetching corpus $\mathcal{C}_i$ in scalable stream, generating prediction $\mathcal{M}_{\mathit{ka}}$ with knowledge acquisition and generating the initial instance $\mathcal{I}$.
We suppose these operations can be finished in $\tilde{O}(1)$.
The initial noisy instance $\mathcal{I}$ is generated based on $\mathcal{M}_{\mathit{ka}}$ i.e.,
\begin{align*}
\mathcal{I} &= \arg\max_{\mathcal{I}_x} \prod_{s_i \in \mathcal{I}_x} \mathcal{M}_{\mathit{ka}}[i][s_i.r] \\
&= \{\langle h_i,r_i,t_i \rangle| p_i = \max_{(r_j, p_j) \in m_i, m_i \in \mathcal{M}_{\mathit{ka}}} p_j \wedge p_i \geq p_{th}\}.
\end{align*}

Secondly, Line 7 involves the validation problem based on graph constraints.
The validation problem is studied in Section \ref{sec:dynamicCommonStructure}, \ref{sec:psRepairValidation} and \ref{sec:graphfraction}.
Section \ref{sec:dynamicCommonStructure} advocates that validation through implicit graph constraints is conditionally equivalent to an explicit case, and utilizing implicit constraints will benefit from an optimizable evaluation $\mathit{Supp}$ for instance validation.
Section \ref{sec:psRepairValidation} proposes to dynamically validate a tuple with implicit graph constraints.
Based on the support set of a localized pattern, we validate tuple $s\in\mathcal{I}$ with $\overline{\mathit{Supp}}'(\mathcal{G}, \mathsf{P}^l_s) \neq \emptyset$, where 
\begin{align*}
\overline{\mathit{Supp}}'(\mathcal{G}, \mathsf{P}^l_s) = \{\mathsf{P} \in S^*(\mathcal{G}, s.r)|\mathsf{sim}(\mathsf{P}^l_s, \mathsf{P}) > 0\}.
\end{align*}

Based on Proposition \ref{lemma:TRcomplexity}, the validation process terminates in $\tilde{O}(\frac{T_{\mathit{valid}}(\mathcal{I})}{|\mathcal{I}|}) = \tilde{O}(\frac{d_M^l*|S^*|}{n}) = \tilde{O}(\frac{d_M^l}{n})$ for each tuple.

Thirdly, Line 8 in Algorithm \ref{algorithm:dynamicStructure} repairs noisy tuples by enumerating possible relation labels.
Relation repair is optimized in Section \ref{sec:cvRelationRepair} and Section \ref{sec:link_prediction}.
Section \ref{sec:cvRelationRepair} proposes an optimize-validate model converting relation repair into instance validation, and it optimizes the instance prediction with linkage prediction.
Section \ref{sec:link_prediction} proposes a linkage prediction model based on graph similarity.
The linkage prediction $\mathcal{M}_{\mathit{lp}}$ is decided by classifier $\mathit{Pr}(r|\langle h,t \rangle, \mathcal{G})$ with 

\begin{align*}
\mathcal{M}_{\mathit{lp}}[i][r] = \mathit{Pr}(r|\langle h_i,t_i \rangle, \mathcal{G}) = \frac{\sum_{\mathsf{P} \in S^*(\mathcal{G}, r)} \mathsf{sim}(\mathsf{P}^l_{\langle h_i,r,t_i \rangle}, \mathsf{P})}{|S^*(\mathcal{G}, r)|}
\end{align*}

The amortized time for relation repair is $\tilde{O}(\frac{k^{|\mathcal{I}|}*T_{\mathit{valid}}(\mathcal{I})+T_{\mathit{lp}}}{|\mathcal{I}|}) = \tilde{O}(\frac{k^{|\mathcal{I}|}*d_M^l}{n})$, which still confronts dynamic hardness.
The hardness comes from enumerating the combination of \rdf tuples with each pair of entities $\langle h,t \rangle$ having $k$ relation labels to vary. 
In this way, combining all possible cases involves $O(k^{|\mathcal{I}|})$ time.
To avoid combination, we repair the instance based on the initial instance $\mathcal{I}$.
When $\mathcal{I}\cup\mathcal{G}$ shows violation, we repair each tuple $t\in\mathcal{I}$ with new relation label taken from \textsc{Top}-k results joint by $\mathcal{M}_{\mathit{ka}}$ and $\mathcal{M}_{\mathit{lp}}$.
Thus, for each tuple $t$, there are $k$ times to change its relation label and pack new instance $\mathcal{I}'=\mathcal{I}\cup\{t'\}\setminus\{s\}$ ; otherwise, the relation label is considered to be \textit{NA}.
Observe this, we have $\tilde{O}(\frac{k*|\mathcal{I}|*T_{\mathit{valid}} + T_{\mathit{lp}}}{|\mathcal{I}|}) = \tilde{O}(\frac{k*d_M^l}{n})$ for relation label repair.

Finally, in Algorithm \ref{algorithm:dynamicStructure}, Line 10 updates graph constraints with new patterns in knowledge graph, which is replaced by the  implicit graph constraints processing.

In conclusion, with the approaches above, we can obtain a dynamic relation repair for knowledge enhancement.
Assuming that $k \ll n$ and $d_M,l$ are constants in one iteration $i$, the overall amortized time consumption $\Delta T_i$ is 

\begin{align*}
\Delta T_i &= \tilde{O}(\frac{T_{\mathit{valid}}(\mathcal{I}_i) + T_{\mathit{repair}}(\mathcal{I}_i) + T_{\mathit{discovery}}(\mathcal{I}_i) }{|\mathcal{I}_i|}) \\
&= \tilde{O}(\frac{d_M^l*\max(|S^*|,k)}{n}) = \tilde{O}(1).
\end{align*}

\subsection{Cold Start Problem}
\label{sec:coldstart}

In this section, we give a brief discussion on the cold start problem during the instance repairing process.
We consider two situations for cold start and provide possible solutions.

\ul{\textbf{Entity Cold Start}.} The size of localized pattern $\mathsf{P}^l_{\langle h,r,t \rangle}$ is rather small e.g., $|\mathsf{P}^l_s|=2$ leads to $|\mathit{TR}(\mathsf{P}^l_s)|=0$ when $l \geq 2$.
Graph matching through embeddings asks for the localized patterns with enough size, otherwise, the embedding will be insufficient.
As the entities $h$ or $t$ are sometimes fresh to the knowledge graph, all knowledge graphs may meet this problem.
For the dynamic situation described in Algorithm \ref{algorithm:dynamicStructure}, the fetched corpus contains a collection of related sentences, which may generate revelant tuples e.g., entity sharing among tuples.
When entity cold starts happen, we try to enhance the knowledge graph after it accumulates enough information i.e., holding the cold start tuples and trying to validate them after a number of iterations.
When some cold-starting tuples still exist, we regard them as an unknown case in Proposition \ref{prop:simplifyVD_tuple}.
These unknown tuples can be added directly or decided by database administrators.


\ul{\textbf{Relation Label Cold Start}.} 
The pattern set $S^*(\mathcal{G}, r)$ is not available or of rather small size.
The poroposed evaluations rely on the localized pattern set $S^*(\mathcal{G}, r)$ to judge the quality of a tuple.
When the pattern set is insufficient, the estimation would be biased.
Since multiple knowledges may share a common subset on relation labels \cite{DBLP:conf/kdd/LeroyCB10}, we use multiple databases to obtain pattern candidates $S^*$.
As graph database integration is time-consuming, we only integrate the relation labels and do not join multiple databases.
Therefore, we formalize this kind of integration as follows.
Given two graph databases $\mathcal{G}=\{V,E,L\}$ and $\mathcal{G}' = \{V',E',L'\}$, we consider a mapping function $\rho$ between two labeling functions $\rho: L'\to L\cup\{(e \to \mathit{NA})\}, \forall e \in \mathcal{G}'.E$, where $\{(e \to \mathit{NA})\}$ is an empty labeling function.
$\rho \circ L'$ marks all edge labels in $\mathcal{G}'$ into the labels in $\mathcal{G}$ and treats the unrelated labels as $\mathit{NA}$.
In this way, more supporting localized patterns could be obtained in $S^*$.

\section{Experiment}
\label{sec:experiment}

In this section, we test our proposed relation repair model over the state-of-the-art knowledge acquisition approaches.
We test the original relation extraction approach and our relation repair approach with datasets \textsc{NYT-FB60K} and FewRel \cite{DBLP:conf/emnlp/GaoHZLLSZ19}, which are both widely used datasets for the relation extraction task.
The test data of \textsc{NYT-FB60K} contains multiple $\mathit{NA}$ tuples, which is different from FewRel.
Multiple graph databases are considered to test the validation efficiency and provide candidates for validation.
We show the detailed dataset information in Table \ref{tab:initial_datasets}.
Our framework is tested under a single node \textsc{Spark} Cluster with 1*Intel(R) i7-8700 and 32 GB RAM.

\begin{table}[t]
  \centering
  \small
  \caption{Datasets information}
  \begin{tabular}{c|c|c|c}
    \toprule
    Graph DB & \#Facts & \#Entities & \#Relations\\
    \midrule
    Wikidata5M & 21M & 3.1M & 822 \\
    YAGO  & 12.4M & 4.29M & 38 \\
    \midrule
    RE Dataset & \#Facts(Train/Test) & \#Entities & \#Relations\\
    \midrule
    NYT-FB60K & 335K/96K & 69K & 1324\\
    FewRel \cite{DBLP:conf/emnlp/GaoHZLLSZ19} & 20.6M/16K & 3.1M & 80\\
    \bottomrule
  \end{tabular}
  \label{tab:initial_datasets}
\end{table}

\begin{table}[t]
  \centering
  \small
  \caption{Initial \textsc{HIT}@1 results and enhanced results}
  \begin{tabular}{c|c|c|c|c}
    \toprule
    Dataset & Model  & Precision & Recall & F-score\\
    \midrule
    \multirow{4}{3.5em}{NYT-FB60K}&JointE \cite{Han2018} & 0.329 & 0.516 & 0.402\\
        &JointE + ImpGFD & 0.807 & 0.564 & 0.664 \\
        &JointD \cite{Han2018} & 0.368 & 0.498 & 0.424 \\
        &JointD + ImpGFD & 0.828 & 0.543 & 0.656 \\
    \hline
    \multirow{2}{3.5em}{FewRel}&CoLAKE \cite{DBLP:conf/coling/SunSQGHHZ20} & 0.903 & 0.902 & 0.903  \\
    &CoLAKE + ImpGFD & 0.932 & 0.872 & 0.901 \\
    \bottomrule
  \end{tabular}
  \label{tab:evalu_init_qe}
\end{table}

\begin{table}[t]
  \centering
  \small
  \caption{Ablation study on our approach}
  \begin{tabular}{c|c|c|c|c}
    \toprule
    Dataset & Model  & Precision & Recall & F-score\\
    \midrule
    \multirow{5}{3.5em}{NYT-FB60K}&JointE & 0.329 & 0.516 & 0.402\\
        &JointE + ImpGFD & 0.807 & 0.564 & 0.664 \\
        &JointE + ImpGFD(LI) & $0.812^*$ & $0.604^*$ & $0.693^*$ \\
        &JointE + LP & 0.629 & 0.572 & 0.599 \\
        &JointE + Valid & 0.820 & 0.543 & 0.651 \\
    \hline
    \multirow{5}{3.5em}{FewRel}&CoLAKE  & 0.903 & 0.902 & 0.903  \\
    &CoLAKE + ImpGFD & 0.932 & 0.872 & 0.901 \\
    &CoLAKE + ImpGFD(LI) & $0.937^*$ & $0.929^*$ & $0.934^*$ \\
    &CoLAKE + LP & 0.899 & 0.899 & 0.899 \\
    &CoLAKE + Valid & 0.938 & 0.863 & 0.898 \\
    \bottomrule
  \end{tabular}
  \label{tab:ablation}
\end{table}

\begin{table}[t]
  \centering
  \small
  \caption{Parameter study on the size of localized patterns}
  \begin{tabular}{c||c|c|c|c}
    \toprule
    Length & Precision & Recall & F-score & Time (MS/Tuple)\\
    \midrule
    $0$ & 0.329 & 0.516 & 0.402 & 0.0\\
    $1$ & 0.764 & 0.519 & 0.618 & 9.7\\
    $2$ & 0.838 & 0.546 & 0.661 & 13.2\\
    $3^*$ & 0.863 & 0.603 & 0.709 & 1034\\
    \bottomrule
  \end{tabular}
  \label{tab:performanceLength}
\end{table}

\subsection{Evaluations for Relation Label Repair}

Three relation extraction models over two datasets are considered to test the efficiency of our repair approach shown in Table \ref{tab:evalu_init_qe}.
We name our relation repair approach as ImpGFD, which utilizes implicitly discovered graph patterns for graph data quality.
The graph database would include an \textsc{RDF} tuple $s$ when $s$ is validated by our dynamic repair approach and the relation label $s.r$ is not $\mathit{NA}$.
The precision score indicates the percentage of true \textsc{RDF} tuples, which reveals the quality of enhanced data and relates to the consistency of the graph database.
The recall and F score evaluates how the model performs on finding new relations from graph structures and corpus.

The precision score in Table \ref{tab:evalu_init_qe} illustrates the consistency of new instances.
The test data of \textsc{NYT-FB60K} contains $\mathit{NA}$ labels, which asks for an RE approach to predict false-negative labels and may affect the efficiency of relation extraction work.
Given each tuple in an instance, our approach would find support for a validation or deletion.
The experimental data on \textsc{NYT-FB60K} indicate that our approach helps to classify false-negative labels by comparing the support set of \textsc{Top}-k relation label predictions of an RE tool; and the data on FewRel dataset indicates that our approach can enhance the graph data consistency without causing many deletions on \rdf tuples.

We provide the ablation study of our approach (ImpGFD) over two relation extraction works on different datasets.
The ablation separates relation repair into two parts i.e., linkage prediction (LP) and validation.
We also provide the quality of legal instances (LI), which is obtained through Proposition \ref{prop:approx_supp_legal}.
To show the quality of legal instances, the scores of LI are evaluated based on non-\textit{NA} tuples while other scores based on the whole instances.
We can conclude that both linkage prediction and instance validation contribute to graph quality enhancement based on Table \ref{tab:ablation}. 
Moreover, a high data quality enhancement to the graph database could be obtained through legal instances, which are validated by non-empty support sets.

\subsection{Evaluation on Dynamic Performance}

\begin{figure*}[t]
  
  \centering
  \hspace{-1em}
  \begin{minipage}{0.24\textwidth}
    \includegraphics[width=0.5\expwidths]{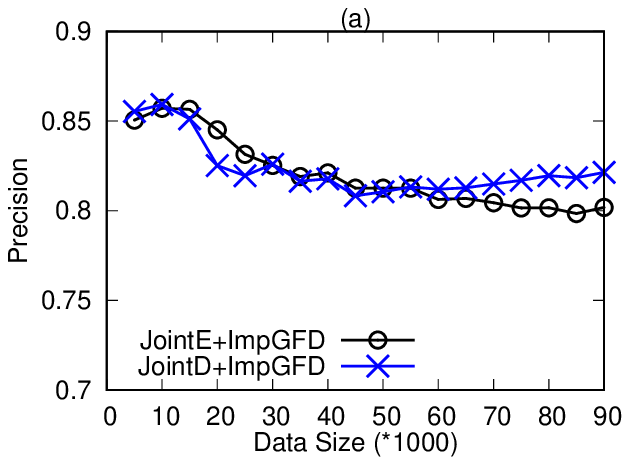}
  \end{minipage}
  \hspace{-1em}%
  \begin{minipage}{0.24\textwidth}
    \includegraphics[width=0.5\expwidths]{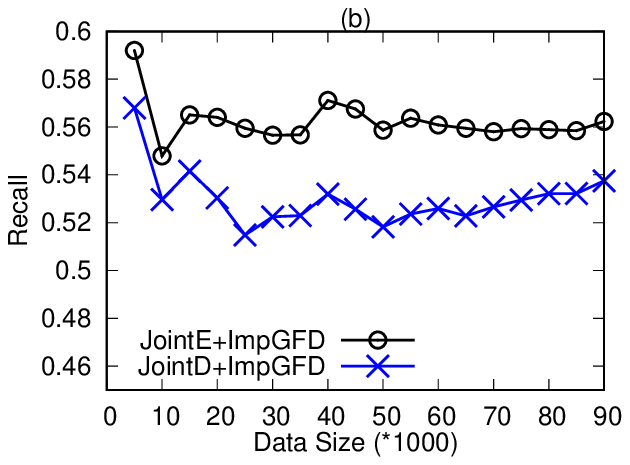}
  \end{minipage}
  \hspace{-1em}
  \begin{minipage}{0.24\textwidth}
    \includegraphics[width=0.5\expwidths]{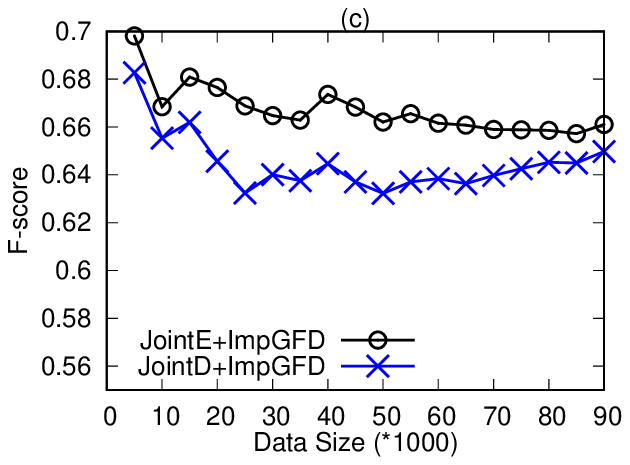}
  \end{minipage}
  \hspace{-1em}
  \begin{minipage}{0.24\textwidth}
    \includegraphics[width=0.5\expwidths]{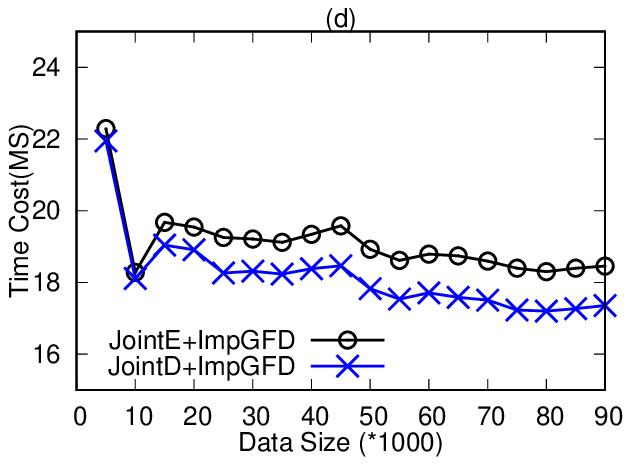}
  \end{minipage}
  \caption{Dynamic performance evaluation}
  \label{fig:dynamic-evaluate}
\end{figure*}

To show dynamic, we introduce this section to test our relation repair approach under scalable data (streamed \textsc{NYT-FB60K}). 
We take both JointE and JointD as knowledge acquisition tools and run a real-time relation extraction based on \textsc{RDF} streams.
We draw the curves on Precision, Recall, F-scores, and time consumption in Figure \ref{fig:dynamic-evaluate}.
Figure \ref{fig:dynamic-evaluate} indicates that the relation repairing approach is stable and dynamic under scalable data.

\subsection{Evaluation on Parameter Study}

\begin{figure*}[t]
	
	\centering
	\hspace{-1em}
	\begin{minipage}{0.24\textwidth}
		\includegraphics[width=0.5\expwidths]{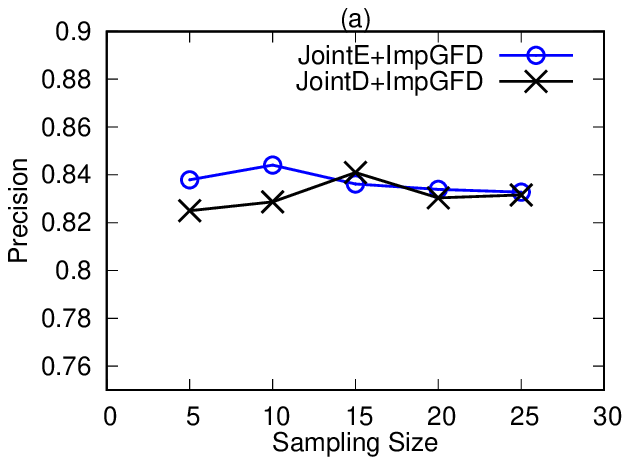}
	\end{minipage}
	\hspace{-1em}%
	\begin{minipage}{0.24\textwidth}
		\includegraphics[width=0.5\expwidths]{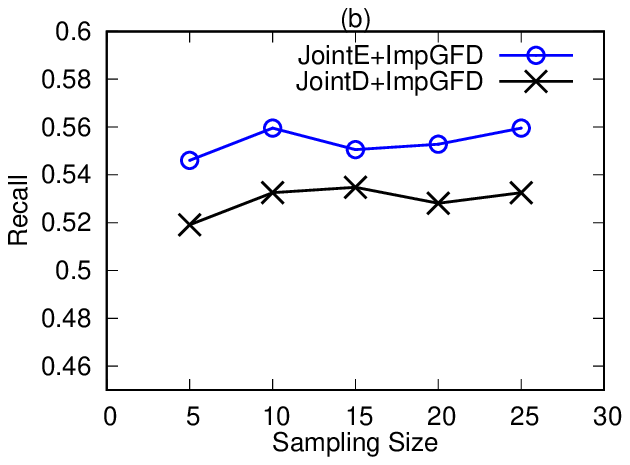}
	\end{minipage}
	\hspace{-1em}
	\begin{minipage}{0.24\textwidth}
		\includegraphics[width=0.5\expwidths]{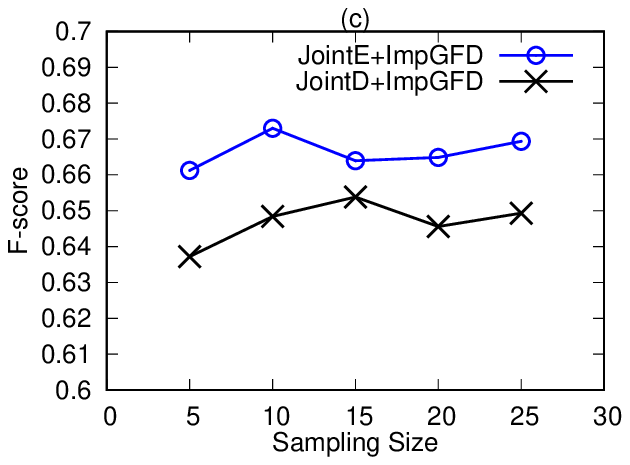}
	\end{minipage}
	\hspace{-1em}
	\begin{minipage}{0.24\textwidth}
		\includegraphics[width=0.5\expwidths]{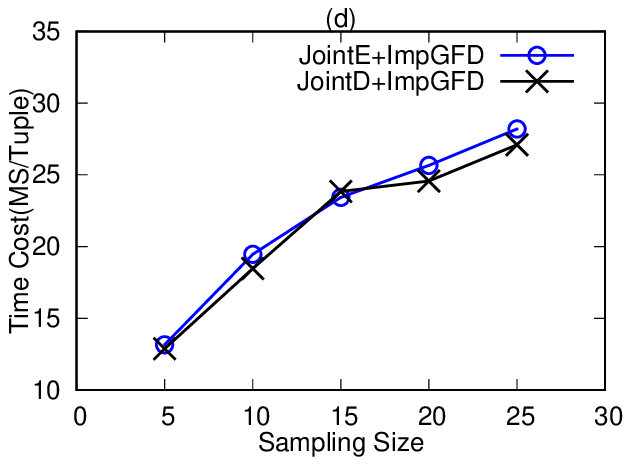}
	\end{minipage}
	\caption{Comparison with various sampling sizes}
	\label{fig:compare-sample-size}
\end{figure*}

To evaluate the effect of different parameters, we use Table \ref{tab:performanceLength} and Figure \ref{fig:compare-sample-size} to discover the performance of relation repairing when varying the size of the localized pattern (decided by $l$) and the size of sampled localized patterns $|S^*|$.
The data in Table \ref{tab:performanceLength} is generated on relation repairing on JointE over \textsc{NYT-FB60K}.
As shown in Table \ref{tab:performanceLength}, when we include more graph structural information by increasing $l$, the ImpGFD framework would provide better performance and consume more time on deciding the relation label of each tuple.
When $|S^*|$ increases, the performance is stable, whereas the time consumption rises as shown in Figure \ref{fig:compare-sample-size}.
Thus, we take $l=2$ and $|S^*|=10$ for the experiments on other studies.

\begin{table}
	\centering
	\small
	\caption{Comparison on error detection}
	\begin{tabular}{c||ccc|ccc}
		\toprule
    Datasets & \multicolumn{3}{c}{YAGO} &  \multicolumn{3}{c}{WikiData} \\
		Approaches & Prec. & Rec. & F1 & Prec. & Rec. & F1\\
		\midrule
		$\mathit{ImpGFD}_{l=1}$ & 0.68 & 0.71 & 0.69 & - & - & -\\
		$\mathit{ImpGFD}_{l=2}$ & 0.72 & \textbf{0.76} & \textbf{0.74} & \textbf{0.83} & 0.64 & \textbf{0.71}\\
		$\mathit{GFact}$\cite{DBLP:journals/jdiq/LinSWP19} & \textbf{0.81} & 0.60 & 0.66 & 0.82 & 0.63 & 0.68\\
		$\mathit{GFact}_R$\cite{DBLP:journals/jdiq/LinSWP19} & 0.40 & 0.75 & 0.50 & 0.55 & 0.64 & 0.55\\
		$\mathit{AMIE}+$\cite{DBLP:journals/vldb/GalarragaTHS15} & 0.44 & 0.76 & 0.51 & 0.42 & \textbf{0.78} & 0.48\\
		$\mathit{PRA}$\cite{Lao2011,Dong2014} & 0.69 & 0.34 & 0.37 & 0.65 & 0.51 & 0.53\\
		$\mathit{KGMiner}$\cite{DBLP:journals/kbs/ShiW16} & 0.62 & 0.36 & 0.40 & 0.63 & 0.49 & 0.52\\
		\bottomrule
	\end{tabular}
	\label{tab:errorDetect}
\end{table}

\subsection{Evaluation for Error Detection}

To examine the performance of our algorithms and avoid the influence of relation extraction tools, we test our approach on the graph error detection task.
Our approach is tested on the efficiency of error detection over \textsc{YAGO} and \textsc{WikiData}, where more relation labels are observed in \textsc{WikiData} based on Table \ref{tab:initial_datasets}.
We sample $80\%$ edges out of full set of facts as training data and the rest $20\%$ facts as test dataset \cite{DBLP:journals/jdiq/LinSWP19}.
In test datasets, we maintain $20\%$ true facts and generate $80\%$ false examples.
Since certain restrictions are applied onto the graph embedding algorithm as shown in Section \ref{sec:graphfraction}, we use Algorithm \ref{algorithm:graphEmbedding} for localized pattern representation learning; otherwise, the Proposition \ref{prop:approx_supp_legal} and \ref{prop:approx_supp_invalid} would degenerate and no longer fit for instance validation.
Table \ref{tab:errorDetect} shows the efficiency of our approach and compares with multiple existing approaches.

Based on Table \ref{tab:errorDetect}, we could notice our algorithms could outperform other works.
Our approach outperforms other works for the following possible reasons.
Firstly, our work would consider more graph patterns as evidence of possible facts including more graph patterns as candidates.
Some relations with fewer occurrences in the knowledge base would not be considered in \cite{DBLP:journals/jdiq/LinSWP19}, however, we keep such patterns serving as evidence.
Apart from this, linkage prediction also predicts possible relation labels and finds potential errors between a pair of entities.


\section{Related Work}
\subsection{Graph Data Repairing}

Our approach focuses on handling the graph relation label quality observed in the process of knowledge enhancement under streamed corpus.
We revise the related works on graph repairing processing, most of which apply support sets or constraints to validate positive patterns for static graph database. 
Graph functional dependency~\cite{DBLP:journals/jdiq/Fan19}, with the form of $\varphi = Q[\overline{x}](X \Rightarrow Y)$, embeds logic expression for the labels or structures in subgraph patterns, and is discovered based on the support of the graph patterns.
Graph constraints processing is composed by three basic problems, validation, satisfaction and implementation~\cite{DBLP:journals/tods/FanL19}, and is proved with hardness on repairing dynamic graph data~\cite{DBLP:conf/sigmod/FanHLL18}.
The validation problem aims to find conflicts within a graph database with respect to a set of predefined graph constraints.
For our approach, we only consider the frequent subgraph patterns with logic clauses on relation labels to simplify the hardness on graph constraint processing.
The approach~\cite{DBLP:conf/www/ChenCHMJ20} also considers to use the support set to build a soft constraint for graph repairing, while it focuses on repairing the entities or literals based on the correctness of relation labels.
The approach~\cite{DBLP:journals/jdiq/LinSWP19} applies support sets of graph patterns with multiple metrics for performance enhancement, and is applicable to static graph validation.
On dealing with the related issues in graph cleaning, \cite{DBLP:conf/www/ZhangDSCZC20}~ focuses on the rare relation label cold start problem on relation label extraction.

\subsection{Knowledge Enhancement}


Knowledge enhancement aims to add new instances to the knowledge graph with consideration to the database consistency.
Knowledge acquisition applies knowledge extraction to three main sources, unstructured data, \textsc{HTML} schemas, and human-annotated pages \cite{Dong2014}.
Existing studies all consider multistage supervision to optimize the performance of knowledge acquisition \cite{Dong2014,Malone2018,DBLP:journals/ijccc/Lin0WWYG18,DBLP:journals/ijswis/PaulheimB14} but few works actually pay attention to validating the acquired knowledge and repairing violations.
As the knowledge acquisition tools generate noisy results \cite{DBLP:journals/ijswis/PaulheimB14}, a direct addition would result in inconsistency in the knowledge database.
Notice that unstructured data is fast-growing and easy to build large knowledge base systems \cite{Dong2014}, approaches to generate new knowledge on the unstructured data stream are recently proposed \cite{DBLP:conf/cikm/000119}.
Therefore, graph data quality processing on an instance stream should be launched.
Linkage prediction aims to catch the missing relations based on knowledge graph\cite{DBLP:conf/nips/Kazemi018,DBLP:conf/cikm/Liben-NowellK03}.
According to its definition, link prediction reconstructs the relation among entities in a knowledge base such as a social graph to solve the incompleteness problem.


\section{Conclusions}
\label{sec:concl}

In this paper, we study the problem of dynamic relation repairing for the knowledge enhancement process.
While enhancing the knowledge graph, we notice that new constraints will emerge for graph data quality.
As it is hard to design graph constraints for graph databases, dynamic constraint discovery is important for graph repair.
We analyze the dynamic hardness of three operations, validation, repair, and discovery based on an algorithmic structure.
We build an optimize-validate model for relation repair with linkage prediction and instance validation.
To eliminate the constraint discovery operation, we prove the equivalence between graph constraint validation and validation through the support subgraph set.
The hardness involved in the cold start problems during knowledge enhancement is further analyzed with applicable solutions.
We prove that with the strategies above, the repairing and enhancing structure is made dynamic.
The experimental result indicates that our approach can aid the sustainability and consistency of the knowledge database.


\bibliographystyle{IEEEtran}
\bibliography{final}

\begin{thebibliography}{10}

\bibitem{DBLP:conf/semweb/AuerBKLCI07}
S.~Auer, C.~Bizer, G.~Kobilarov, J.~Lehmann, R.~Cyganiak, and Z.~G. Ives.
\newblock Dbpedia: {A} nucleus for a web of open data.
\newblock In K.~Aberer, K.~Choi, N.~F. Noy, D.~Allemang, K.~Lee, L.~J.~B.
  Nixon, J.~Golbeck, P.~Mika, D.~Maynard, R.~Mizoguchi, G.~Schreiber, and
  P.~Cudr{\'{e}}{-}Mauroux, editors, {\em The Semantic Web, 6th International
  Semantic Web Conference, 2nd Asian Semantic Web Conference, {ISWC} 2007 +
  {ASWC} 2007, Busan, Korea, November 11-15, 2007}, volume 4825 of {\em Lecture
  Notes in Computer Science}, pages 722--735. Springer, 2007.

\bibitem{Bollacker2008}
K.~D. Bollacker, C.~Evans, P.~Paritosh, T.~Sturge, and J.~Taylor.
\newblock Freebase: a collaboratively created graph database for structuring
  human knowledge.
\newblock In J.~T. Wang, editor, {\em Proceedings of the {ACM} {SIGMOD}
  International Conference on Management of Data, {SIGMOD} 2008, Vancouver, BC,
  Canada, June 10-12, 2008}, pages 1247--1250. {ACM}, 2008.

\bibitem{DBLP:journals/corr/abs-1909-00958}
F.~Chen, Y.~Wang, B.~Wang, and C.~J. Kuo.
\newblock Graph representation learning: {A} survey.
\newblock {\em CoRR}, abs/1909.00958, 2019.

\bibitem{DBLP:journals/corr/abs-2004-11861}
T.~S. Costa, S.~Gottschalk, and E.~Demidova.
\newblock Event-qa: {A} dataset for event-centric question answering over
  knowledge graphs.
\newblock {\em CoRR}, abs/2004.11861, 2020.

\bibitem{DBLP:journals/corr/abs-1903-02419}
W.~Cui, Y.~Xiao, H.~Wang, Y.~Song, S.~Hwang, and W.~Wang.
\newblock {KBQA:} learning question answering over {QA} corpora and knowledge
  bases.
\newblock {\em CoRR}, abs/1903.02419, 2019.

\bibitem{Dong2014}
X.~Dong, E.~Gabrilovich, G.~Heitz, W.~Horn, N.~Lao, K.~Murphy, T.~Strohmann,
  S.~Sun, and W.~Zhang.
\newblock Knowledge vault: a web-scale approach to probabilistic knowledge
  fusion.
\newblock In S.~A. Macskassy, C.~Perlich, J.~Leskovec, W.~Wang, and R.~Ghani,
  editors, {\em The 20th {ACM} {SIGKDD} International Conference on Knowledge
  Discovery and Data Mining, {KDD} '14, New York, NY, {USA} - August 24 - 27,
  2014}, pages 601--610. {ACM}, 2014.

\bibitem{DBLP:journals/jdiq/Fan19}
W.~Fan.
\newblock Dependencies for graphs: Challenges and opportunities.
\newblock {\em J. Data and Information Quality}, 11(2):5:1--5:12, 2019.

\bibitem{DBLP:conf/sigmod/FanHLL18}
W.~Fan, C.~Hu, X.~Liu, and P.~Lu.
\newblock Discovering graph functional dependencies.
\newblock In {\em Proceedings of the 2018 International Conference on
  Management of Data, {SIGMOD} Conference 2018, Houston, TX, USA, June 10-15,
  2018}, pages 427--439, 2018.

\bibitem{DBLP:journals/tods/FanL19}
W.~Fan and P.~Lu.
\newblock Dependencies for graphs.
\newblock {\em {ACM} Trans. Database Syst.}, 44(2):5:1--5:40, 2019.

\bibitem{DBLP:journals/vldb/GalarragaTHS15}
L.~Gal{\'{a}}rraga, C.~Teflioudi, K.~Hose, and F.~M. Suchanek.
\newblock Fast rule mining in ontological knowledge bases with {AMIE+}.
\newblock {\em {VLDB} J.}, 24(6):707--730, 2015.

\bibitem{DBLP:conf/cikm/000119}
J.~Han.
\newblock From unstructured text to textcube: Automated construction and
  multidimensional exploration.
\newblock In W.~Zhu, D.~Tao, X.~Cheng, P.~Cui, E.~A. Rundensteiner, D.~Carmel,
  Q.~He, and J.~X. Yu, editors, {\em Proceedings of the 28th {ACM}
  International Conference on Information and Knowledge Management, {CIKM}
  2019, Beijing, China, November 3-7, 2019}, pages 5--6. {ACM}, 2019.

\bibitem{Han2018}
X.~Han, Z.~Liu, and M.~Sun.
\newblock Neural knowledge acquisition via mutual attention between knowledge
  graph and text.
\newblock In S.~A. McIlraith and K.~Q. Weinberger, editors, {\em Proceedings of
  the Thirty-Second {AAAI} Conference on Artificial Intelligence, (AAAI-18),
  the 30th innovative Applications of Artificial Intelligence (IAAI-18), and
  the 8th {AAAI} Symposium on Educational Advances in Artificial Intelligence
  (EAAI-18), New Orleans, Louisiana, USA, February 2-7, 2018}, pages
  4832--4839. {AAAI} Press, 2018.

\bibitem{DBLP:conf/www/HoffartSBLMW11}
J.~Hoffart, F.~M. Suchanek, K.~Berberich, E.~Lewis{-}Kelham, G.~de~Melo, and
  G.~Weikum.
\newblock {YAGO2:} exploring and querying world knowledge in time, space,
  context, and many languages.
\newblock In S.~Srinivasan, K.~Ramamritham, A.~Kumar, M.~P. Ravindra,
  E.~Bertino, and R.~Kumar, editors, {\em Proceedings of the 20th International
  Conference on World Wide Web, {WWW} 2011, Hyderabad, India, March 28 - April
  1, 2011 (Companion Volume)}, pages 229--232. {ACM}, 2011.

\bibitem{DBLP:journals/corr/abs-2002-00388}
S.~Ji, S.~Pan, E.~Cambria, P.~Marttinen, and P.~S. Yu.
\newblock A survey on knowledge graphs: Representation, acquisition and
  applications.
\newblock {\em CoRR}, abs/2002.00388, 2020.

\bibitem{DBLP:journals/ker/JiangCZ13}
C.~Jiang, F.~Coenen, and M.~Zito.
\newblock A survey of frequent subgraph mining algorithms.
\newblock {\em Knowledge Eng. Review}, 28(1):75--105, 2013.

\bibitem{DBLP:conf/nips/Kazemi018}
S.~M. Kazemi and D.~Poole.
\newblock Simple embedding for link prediction in knowledge graphs.
\newblock In S.~Bengio, H.~M. Wallach, H.~Larochelle, K.~Grauman,
  N.~Cesa{-}Bianchi, and R.~Garnett, editors, {\em Advances in Neural
  Information Processing Systems 31: Annual Conference on Neural Information
  Processing Systems 2018, NeurIPS 2018, 3-8 December 2018, Montr{\'{e}}al,
  Canada}, pages 4289--4300, 2018.

\bibitem{Lao2011}
N.~Lao, T.~M. Mitchell, and W.~W. Cohen.
\newblock Random walk inference and learning in {A} large scale knowledge base.
\newblock In {\em Proceedings of the 2011 Conference on Empirical Methods in
  Natural Language Processing, {EMNLP} 2011, 27-31 July 2011, John McIntyre
  Conference Centre, Edinburgh, UK, {A} meeting of SIGDAT, a Special Interest
  Group of the {ACL}}, pages 529--539. {ACL}, 2011.

\bibitem{DBLP:conf/kdd/LeroyCB10}
V.~Leroy, B.~B. Cambazoglu, and F.~Bonchi.
\newblock Cold start link prediction.
\newblock In B.~Rao, B.~Krishnapuram, A.~Tomkins, and Q.~Yang, editors, {\em
  Proceedings of the 16th {ACM} {SIGKDD} International Conference on Knowledge
  Discovery and Data Mining, Washington, DC, USA, July 25-28, 2010}, pages
  393--402. {ACM}, 2010.

\bibitem{DBLP:conf/cikm/Liben-NowellK03}
D.~Liben{-}Nowell and J.~M. Kleinberg.
\newblock The link prediction problem for social networks.
\newblock In {\em Proceedings of the 2003 {ACM} {CIKM} International Conference
  on Information and Knowledge Management, New Orleans, Louisiana, USA,
  November 2-8, 2003}, pages 556--559. {ACM}, 2003.

\bibitem{DBLP:journals/jdiq/LinSWP19}
P.~Lin, Q.~Song, Y.~Wu, and J.~Pi.
\newblock Discovering patterns for fact checking in knowledge graphs.
\newblock {\em J. Data and Information Quality}, 11(3):13:1--13:27, 2019.

\bibitem{DBLP:journals/ijccc/Lin0WWYG18}
X.~Lin, Y.~Liang, L.~Wang, X.~Wang, M.~Q. Yang, and R.~Guan.
\newblock A knowledge base completion model based on path feature learning.
\newblock {\em Int. J. Comput. Commun. Control}, 13(1):71--82, 2018.

\bibitem{Malone2018}
B.~Malone, A.~Garc{\'{\i}}a{-}Dur{\'{a}}n, and M.~Niepert.
\newblock Knowledge graph completion to predict polypharmacy side effects.
\newblock {\em CoRR}, abs/1810.09227, 2018.

\bibitem{DBLP:journals/ijswis/PaulheimB14}
H.~Paulheim and C.~Bizer.
\newblock Improving the quality of linked data using statistical distributions.
\newblock {\em Int. J. Semantic Web Inf. Syst.}, 10(2):63--86, 2014.

\bibitem{DBLP:journals/jiis/ProtaziukLB16}
G.~Protaziuk, J.~Lewandowski, and R.~Bembenik.
\newblock Sautext - a system for analysis of unstructured textual data.
\newblock {\em J. Intell. Inf. Syst.}, 46(2):369--389, 2016.

\bibitem{Riedel2013}
S.~Riedel, L.~Yao, A.~McCallum, and B.~M. Marlin.
\newblock Relation extraction with matrix factorization and universal schemas.
\newblock In L.~Vanderwende, H.~D. III, and K.~Kirchhoff, editors, {\em Human
  Language Technologies: Conference of the North American Chapter of the
  Association of Computational Linguistics, Proceedings, June 9-14, 2013,
  Westin Peachtree Plaza Hotel, Atlanta, Georgia, {USA}}, pages 74--84. The
  Association for Computational Linguistics, 2013.

\bibitem{DBLP:journals/corr/abs-2002-00819}
A.~Rossi, D.~Firmani, A.~Matinata, P.~Merialdo, and D.~Barbosa.
\newblock Knowledge graph embedding for link prediction: {A} comparative
  analysis.
\newblock {\em CoRR}, abs/2002.00819, 2020.

\bibitem{DBLP:journals/kbs/ShiW16}
B.~Shi and T.~Weninger.
\newblock Discriminative predicate path mining for fact checking in knowledge
  graphs.
\newblock {\em Knowl. Based Syst.}, 104:123--133, 2016.

\bibitem{DBLP:journals/tods/Song011}
S.~Song and L.~Chen.
\newblock Differential dependencies: Reasoning and discovery.
\newblock {\em {ACM} Trans. Database Syst.}, 36(3):16:1--16:41, 2011.

\bibitem{DBLP:conf/wikis/SuhCCP09}
B.~Suh, G.~Convertino, E.~H. Chi, and P.~Pirolli.
\newblock The singularity is not near: slowing growth of wikipedia.
\newblock In D.~Riehle and A.~Bruckman, editors, {\em Proceedings of the 2009
  International Symposium on Wikis, 2009, Orlando, Florida, USA, October 25-27,
  2009}. {ACM}, 2009.

\bibitem{DBLP:journals/corr/abs-1911-00219}
S.~Vashishth, S.~Sanyal, V.~Nitin, N.~Agrawal, and P.~P. Talukdar.
\newblock Interacte: Improving convolution-based knowledge graph embeddings by
  increasing feature interactions.
\newblock {\em CoRR}, abs/1911.00219, 2019.

\bibitem{Weikum2010}
G.~Weikum and M.~Theobald.
\newblock From information to knowledge: harvesting entities and relationships
  from web sources.
\newblock In J.~Paredaens and D.~V. Gucht, editors, {\em Proceedings of the
  Twenty-Ninth {ACM} {SIGMOD-SIGACT-SIGART} Symposium on Principles of Database
  Systems, {PODS} 2010, June 6-11, 2010, Indianapolis, Indiana, {USA}}, pages
  65--76. {ACM}, 2010.

\bibitem{DBLP:journals/corr/abs-1901-00596}
Z.~Wu, S.~Pan, F.~Chen, G.~Long, C.~Zhang, and P.~S. Yu.
\newblock A comprehensive survey on graph neural networks.
\newblock {\em CoRR}, abs/1901.00596, 2019.

\bibitem{DBLP:journals/corr/abs-1906-00687}
C.~Xu and R.~Li.
\newblock Relation embedding with dihedral group in knowledge graph.
\newblock {\em CoRR}, abs/1906.00687, 2019.

\bibitem{Zeng2015}
D.~Zeng, K.~Liu, Y.~Chen, and J.~Zhao.
\newblock Distant supervision for relation extraction via piecewise
  convolutional neural networks.
\newblock In L.~M{\`{a}}rquez, C.~Callison{-}Burch, J.~Su, D.~Pighin, and
  Y.~Marton, editors, {\em Proceedings of the 2015 Conference on Empirical
  Methods in Natural Language Processing, {EMNLP} 2015, Lisbon, Portugal,
  September 17-21, 2015}, pages 1753--1762. The Association for Computational
  Linguistics, 2015.

\bibitem{Zheng2018}
W.~Zheng, J.~X. Yu, L.~Zou, and H.~Cheng.
\newblock Question answering over knowledge graphs: Question understanding via
  template decomposition.
\newblock {\em {PVLDB}}, 11(11):1373--1386, 2018.

\bibitem{DBLP:journals/corr/abs-1902-00756}
H.~Zhu, Y.~Lin, Z.~Liu, J.~Fu, T.~Chua, and M.~Sun.
\newblock Graph neural networks with generated parameters for relation
  extraction.
\newblock {\em CoRR}, abs/1902.00756, 2019.

\end{thebibliography}


\begin{thebibliography}{10}
\providecommand{\url}[1]{#1}
\csname url@samestyle\endcsname
\providecommand{\newblock}{\relax}
\providecommand{\bibinfo}[2]{#2}
\providecommand{\BIBentrySTDinterwordspacing}{\spaceskip=0pt\relax}
\providecommand{\BIBentryALTinterwordstretchfactor}{4}
\providecommand{\BIBentryALTinterwordspacing}{\spaceskip=\fontdimen2\font plus
\BIBentryALTinterwordstretchfactor\fontdimen3\font minus
  \fontdimen4\font\relax}
\providecommand{\BIBforeignlanguage}[2]{{%
\expandafter\ifx\csname l@#1\endcsname\relax
\typeout{** WARNING: IEEEtran.bst: No hyphenation pattern has been}%
\typeout{** loaded for the language `#1'. Using the pattern for}%
\typeout{** the default language instead.}%
\else
\language=\csname l@#1\endcsname
\fi
#2}}
\providecommand{\BIBdecl}{\relax}
\BIBdecl

\bibitem{DBLP:journals/corr/abs-2004-11861}
T.~S. Costa, S.~Gottschalk, and E.~Demidova, ``Event-qa: {A} dataset for
  event-centric question answering over knowledge graphs,'' \emph{CoRR}, 2020.

\bibitem{DBLP:journals/corr/abs-1903-02419}
W.~Cui, Y.~Xiao, H.~Wang, Y.~Song, S.~Hwang, and W.~Wang, ``{KBQA:} learning
  question answering over {QA} corpora and knowledge bases,'' \emph{CoRR},
  2019.

\bibitem{Zheng2018}
W.~Zheng, J.~X. Yu, L.~Zou, and H.~Cheng, ``Question answering over knowledge
  graphs: Question understanding via template decomposition,'' \emph{{PVLDB}},
  vol.~11, no.~11, pp. 1373--1386, 2018.

\bibitem{DBLP:conf/www/HoffartSBLMW11}
J.~Hoffart, F.~M. Suchanek, K.~Berberich, E.~Lewis{-}Kelham, G.~de~Melo, and
  G.~Weikum, ``{YAGO2:} exploring and querying world knowledge in time, space,
  context, and many languages,'' in \emph{Proceedings of the 20th International
  Conference on World Wide Web, {WWW} 2011, Hyderabad, India, March 28 - April
  1, 2011 (Companion Volume)}, S.~Srinivasan, K.~Ramamritham, A.~Kumar, M.~P.
  Ravindra, E.~Bertino, and R.~Kumar, Eds.\hskip 1em plus 0.5em minus
  0.4em\relax {ACM}, 2011, pp. 229--232.

\bibitem{Bollacker2008}
K.~D. Bollacker, C.~Evans, P.~Paritosh, T.~Sturge, and J.~Taylor, ``Freebase: a
  collaboratively created graph database for structuring human knowledge,'' in
  \emph{Proceedings of the {ACM} {SIGMOD} International Conference on
  Management of Data, {SIGMOD} 2008, Vancouver, BC, Canada, June 10-12, 2008},
  J.~T. Wang, Ed.\hskip 1em plus 0.5em minus 0.4em\relax {ACM}, 2008, pp.
  1247--1250.

\bibitem{Dong2014}
X.~Dong, E.~Gabrilovich, G.~Heitz, W.~Horn, N.~Lao, K.~Murphy, T.~Strohmann,
  S.~Sun, and W.~Zhang, ``Knowledge vault: a web-scale approach to
  probabilistic knowledge fusion,'' in \emph{The 20th {ACM} {SIGKDD}
  International Conference on Knowledge Discovery and Data Mining, {KDD} '14,
  New York, NY, {USA} - August 24 - 27, 2014}, S.~A. Macskassy, C.~Perlich,
  J.~Leskovec, W.~Wang, and R.~Ghani, Eds.\hskip 1em plus 0.5em minus
  0.4em\relax {ACM}, 2014, pp. 601--610.

\bibitem{DBLP:conf/wikis/SuhCCP09}
B.~Suh, G.~Convertino, E.~H. Chi, and P.~Pirolli, ``The singularity is not
  near: slowing growth of wikipedia,'' in \emph{Proceedings of the 2009
  International Symposium on Wikis, 2009, Orlando, Florida, USA, October 25-27,
  2009}, D.~Riehle and A.~Bruckman, Eds.\hskip 1em plus 0.5em minus 0.4em\relax
  {ACM}, 2009.

\bibitem{DBLP:journals/jiis/ProtaziukLB16}
G.~Protaziuk, J.~Lewandowski, and R.~Bembenik, ``Sautext - a system for
  analysis of unstructured textual data,'' \emph{J. Intell. Inf. Syst.},
  vol.~46, no.~2, pp. 369--389, 2016.

\bibitem{Zeng2015}
D.~Zeng, K.~Liu, Y.~Chen, and J.~Zhao, ``Distant supervision for relation
  extraction via piecewise convolutional neural networks,'' in
  \emph{Proceedings of the 2015 Conference on Empirical Methods in Natural
  Language Processing, {EMNLP} 2015, Lisbon, Portugal, September 17-21, 2015},
  L.~M{\`{a}}rquez, C.~Callison{-}Burch, J.~Su, D.~Pighin, and Y.~Marton,
  Eds.\hskip 1em plus 0.5em minus 0.4em\relax The Association for Computational
  Linguistics, 2015, pp. 1753--1762.

\bibitem{Han2018}
X.~Han, Z.~Liu, and M.~Sun, ``Neural knowledge acquisition via mutual attention
  between knowledge graph and text,'' in \emph{Proceedings of the Thirty-Second
  {AAAI} Conference on Artificial Intelligence, (AAAI-18), the 30th innovative
  Applications of Artificial Intelligence (IAAI-18), and the 8th {AAAI}
  Symposium on Educational Advances in Artificial Intelligence (EAAI-18), New
  Orleans, Louisiana, USA, February 2-7, 2018}, S.~A. McIlraith and K.~Q.
  Weinberger, Eds.\hskip 1em plus 0.5em minus 0.4em\relax {AAAI} Press, 2018,
  pp. 4832--4839.

\bibitem{DBLP:journals/corr/abs-1902-00756}
H.~Zhu, Y.~Lin, Z.~Liu, J.~Fu, T.~Chua, and M.~Sun, ``Graph neural networks
  with generated parameters for relation extraction,'' \emph{CoRR}, 2019.

\bibitem{DBLP:journals/corr/abs-1906-00687}
C.~Xu and R.~Li, ``Relation embedding with dihedral group in knowledge graph,''
  \emph{CoRR}, vol. abs/1906.00687, 2019.

\bibitem{DBLP:journals/corr/abs-1911-00219}
S.~Vashishth, S.~Sanyal, V.~Nitin, N.~Agrawal, and P.~P. Talukdar, ``Interacte:
  Improving convolution-based knowledge graph embeddings by increasing feature
  interactions,'' \emph{CoRR}, 2019.

\bibitem{Riedel2013}
S.~Riedel, L.~Yao, A.~McCallum, and B.~M. Marlin, ``Relation extraction with
  matrix factorization and universal schemas,'' in \emph{Human Language
  Technologies: Conference of the North American Chapter of the Association of
  Computational Linguistics, Proceedings, June 9-14, 2013, Westin Peachtree
  Plaza Hotel, Atlanta, Georgia, {USA}}, L.~Vanderwende, H.~D. III, and
  K.~Kirchhoff, Eds.\hskip 1em plus 0.5em minus 0.4em\relax The Association for
  Computational Linguistics, 2013, pp. 74--84.

\bibitem{Weikum2010}
G.~Weikum and M.~Theobald, ``From information to knowledge: harvesting entities
  and relationships from web sources,'' in \emph{Proceedings of the
  Twenty-Ninth {ACM} {SIGMOD-SIGACT-SIGART} Symposium on Principles of Database
  Systems, {PODS} 2010, June 6-11, 2010, Indianapolis, Indiana, {USA}},
  J.~Paredaens and D.~V. Gucht, Eds.\hskip 1em plus 0.5em minus 0.4em\relax
  {ACM}, 2010, pp. 65--76.

\bibitem{DBLP:journals/jdiq/Fan19}
W.~Fan, ``Dependencies for graphs: Challenges and opportunities,'' \emph{J.
  Data and Information Quality}, vol.~11, no.~2, pp. 5:1--5:12, 2019.

\bibitem{DBLP:conf/sigmod/FanHLL18}
W.~Fan, C.~Hu, X.~Liu, and P.~Lu, ``Discovering graph functional
  dependencies,'' in \emph{Proceedings of the 2018 International Conference on
  Management of Data, {SIGMOD} Conference 2018, Houston, TX, USA, June 10-15,
  2018}, 2018, pp. 427--439.

\bibitem{DBLP:journals/tods/Song011}
S.~Song and L.~Chen, ``Differential dependencies: Reasoning and discovery,''
  \emph{{ACM} Trans. Database Syst.}, vol.~36, no.~3, pp. 16:1--16:41, 2011.

\bibitem{DBLP:journals/ijccc/Lin0WWYG18}
X.~Lin, Y.~Liang, L.~Wang, X.~Wang, M.~Q. Yang, and R.~Guan, ``A knowledge base
  completion model based on path feature learning,'' \emph{Int. J. Comput.
  Commun. Control}, vol.~13, no.~1, pp. 71--82, 2018.

\bibitem{DBLP:journals/ijswis/PaulheimB14}
H.~Paulheim and C.~Bizer, ``Improving the quality of linked data using
  statistical distributions,'' \emph{Int. J. Semantic Web Inf. Syst.}, vol.~10,
  no.~2, pp. 63--86, 2014.

\bibitem{DBLP:conf/icde/0001SLZP15}
J.~Wang, S.~Song, X.~Lin, X.~Zhu, and J.~Pei, ``Cleaning structured event logs:
  {A} graph repair approach,'' in \emph{31st {IEEE} International Conference on
  Data Engineering, {ICDE} 2015, Seoul, South Korea, April 13-17, 2015},
  J.~Gehrke, W.~Lehner, K.~Shim, S.~K. Cha, and G.~M. Lohman, Eds.\hskip 1em
  plus 0.5em minus 0.4em\relax {IEEE} Computer Society, 2015, pp. 30--41.

\bibitem{DBLP:journals/tods/FanL19}
W.~Fan and P.~Lu, ``Dependencies for graphs,'' \emph{{ACM} Trans. Database
  Syst.}, vol.~44, no.~2, pp. 5:1--5:40, 2019.

\bibitem{DBLP:conf/amw/Corte-CalabuigP12}
A.~Cort{\'{e}}s{-}Calabuig and J.~Paredaens, ``Semantics of constraints in
  {RDFS},'' in \emph{Proceedings of the 6th Alberto Mendelzon International
  Workshop on Foundations of Data Management, Ouro Preto, Brazil, June 27-30,
  2012}, ser. {CEUR} Workshop Proceedings, J.~Freire and D.~Suciu, Eds., vol.
  866.\hskip 1em plus 0.5em minus 0.4em\relax CEUR-WS.org, 2012, pp. 75--90.

\bibitem{DBLP:journals/corr/abs-2002-00819}
A.~Rossi, D.~Firmani, A.~Matinata, P.~Merialdo, and D.~Barbosa, ``Knowledge
  graph embedding for link prediction: {A} comparative analysis,'' \emph{CoRR},
  2020.

\bibitem{DBLP:conf/cikm/Liben-NowellK03}
D.~Liben{-}Nowell and J.~M. Kleinberg, ``The link prediction problem for social
  networks,'' in \emph{Proceedings of the 2003 {ACM} {CIKM} International
  Conference on Information and Knowledge Management, New Orleans, Louisiana,
  USA, November 2-8, 2003}.\hskip 1em plus 0.5em minus 0.4em\relax {ACM}, 2003,
  pp. 556--559.

\bibitem{DBLP:conf/nips/Kazemi018}
S.~M. Kazemi and D.~Poole, ``Simple embedding for link prediction in knowledge
  graphs,'' in \emph{Advances in Neural Information Processing Systems 31:
  Annual Conference on Neural Information Processing Systems 2018, NeurIPS
  2018, 3-8 December 2018, Montr{\'{e}}al, Canada}, S.~Bengio, H.~M. Wallach,
  H.~Larochelle, K.~Grauman, N.~Cesa{-}Bianchi, and R.~Garnett, Eds., 2018, pp.
  4289--4300.

\bibitem{DBLP:conf/kdd/LeroyCB10}
V.~Leroy, B.~B. Cambazoglu, and F.~Bonchi, ``Cold start link prediction,'' in
  \emph{Proceedings of the 16th {ACM} {SIGKDD} International Conference on
  Knowledge Discovery and Data Mining, Washington, DC, USA, July 25-28, 2010},
  B.~Rao, B.~Krishnapuram, A.~Tomkins, and Q.~Yang, Eds.\hskip 1em plus 0.5em
  minus 0.4em\relax {ACM}, 2010, pp. 393--402.

\bibitem{DBLP:journals/corr/abs-1909-00958}
F.~Chen, Y.~Wang, B.~Wang, and C.~J. Kuo, ``Graph representation learning: {A}
  survey,'' \emph{CoRR}, vol. abs/1909.00958, 2019.

\bibitem{DBLP:journals/corr/abs-2002-00388}
S.~Ji, S.~Pan, E.~Cambria, P.~Marttinen, and P.~S. Yu, ``A survey on knowledge
  graphs: Representation, acquisition and applications,'' \emph{CoRR}, vol.
  abs/2002.00388, 2020.

\bibitem{DBLP:journals/jmlr/KazemiGJKSFP20}
S.~M. Kazemi, R.~Goel, K.~Jain, I.~Kobyzev, A.~Sethi, P.~Forsyth, and
  P.~Poupart, ``Representation learning for dynamic graphs: {A} survey,''
  \emph{J. Mach. Learn. Res.}, vol.~21, pp. 70:1--70:73, 2020.

\bibitem{DBLP:conf/emnlp/GaoHZLLSZ19}
T.~Gao, X.~Han, H.~Zhu, Z.~Liu, P.~Li, M.~Sun, and J.~Zhou, ``Fewrel 2.0:
  Towards more challenging few-shot relation classification,'' in
  \emph{Proceedings of the 2019 Conference on Empirical Methods in Natural
  Language Processing and the 9th International Joint Conference on Natural
  Language Processing, {EMNLP-IJCNLP} 2019, Hong Kong, China, November 3-7,
  2019}, K.~Inui, J.~Jiang, V.~Ng, and X.~Wan, Eds.\hskip 1em plus 0.5em minus
  0.4em\relax Association for Computational Linguistics, 2019, pp. 6249--6254.

\bibitem{DBLP:conf/coling/SunSQGHHZ20}
T.~Sun, Y.~Shao, X.~Qiu, Q.~Guo, Y.~Hu, X.~Huang, and Z.~Zhang, ``Colake:
  Contextualized language and knowledge embedding,'' in \emph{Proceedings of
  the 28th International Conference on Computational Linguistics, {COLING}
  2020, Barcelona, Spain (Online), December 8-13, 2020}, D.~Scott, N.~Bel, and
  C.~Zong, Eds.\hskip 1em plus 0.5em minus 0.4em\relax International Committee
  on Computational Linguistics, 2020, pp. 3660--3670.

\bibitem{DBLP:journals/jdiq/LinSWP19}
P.~Lin, Q.~Song, Y.~Wu, and J.~Pi, ``Discovering patterns for fact checking in
  knowledge graphs,'' \emph{J. Data and Information Quality}, vol.~11, no.~3,
  pp. 13:1--13:27, 2019.

\bibitem{DBLP:journals/vldb/GalarragaTHS15}
L.~Gal{\'{a}}rraga, C.~Teflioudi, K.~Hose, and F.~M. Suchanek, ``Fast rule
  mining in ontological knowledge bases with {AMIE+},'' \emph{{VLDB} J.},
  vol.~24, no.~6, pp. 707--730, 2015.

\bibitem{Lao2011}
N.~Lao, T.~M. Mitchell, and W.~W. Cohen, ``Random walk inference and learning
  in {A} large scale knowledge base,'' in \emph{Proceedings of the 2011
  Conference on Empirical Methods in Natural Language Processing, {EMNLP} 2011,
  27-31 July 2011, John McIntyre Conference Centre, Edinburgh, UK, {A} meeting
  of SIGDAT, a Special Interest Group of the {ACL}}.\hskip 1em plus 0.5em minus
  0.4em\relax {ACL}, 2011, pp. 529--539.

\bibitem{DBLP:journals/kbs/ShiW16}
B.~Shi and T.~Weninger, ``Discriminative predicate path mining for fact
  checking in knowledge graphs,'' \emph{Knowl. Based Syst.}, vol. 104, pp.
  123--133, 2016.

\bibitem{DBLP:conf/www/ChenCHMJ20}
\BIBentryALTinterwordspacing
J.~Chen, X.~Chen, I.~Horrocks, E.~B. Myklebust, and E.~Jim{\'{e}}nez{-}Ruiz,
  ``Correcting knowledge base assertions,'' in \emph{{WWW} '20: The Web
  Conference 2020, Taipei, Taiwan, April 20-24, 2020}, Y.~Huang, I.~King,
  T.~Liu, and M.~van Steen, Eds.\hskip 1em plus 0.5em minus 0.4em\relax {ACM} /
  {IW3C2}, 2020, pp. 1537--1547. [Online]. Available:
  \url{https://doi.org/10.1145/3366423.3380226}
\BIBentrySTDinterwordspacing

\bibitem{DBLP:conf/www/ZhangDSCZC20}
\BIBentryALTinterwordspacing
N.~Zhang, S.~Deng, Z.~Sun, J.~Chen, W.~Zhang, and H.~Chen, ``Relation
  adversarial network for low resource knowledge graph completion,'' in
  \emph{{WWW} '20: The Web Conference 2020, Taipei, Taiwan, April 20-24, 2020},
  Y.~Huang, I.~King, T.~Liu, and M.~van Steen, Eds.\hskip 1em plus 0.5em minus
  0.4em\relax {ACM} / {IW3C2}, 2020, pp. 1--12. [Online]. Available:
  \url{https://doi.org/10.1145/3366423.3380089}
\BIBentrySTDinterwordspacing

\bibitem{Malone2018}
B.~Malone, A.~Garc{\'{\i}}a{-}Dur{\'{a}}n, and M.~Niepert, ``Knowledge graph
  completion to predict polypharmacy side effects,'' \emph{CoRR}, vol.
  abs/1810.09227, 2018.

\bibitem{DBLP:conf/cikm/000119}
J.~Han, ``From unstructured text to textcube: Automated construction and
  multidimensional exploration,'' in \emph{Proceedings of the 28th {ACM}
  International Conference on Information and Knowledge Management, {CIKM}
  2019, Beijing, China, November 3-7, 2019}, W.~Zhu, D.~Tao, X.~Cheng, P.~Cui,
  E.~A. Rundensteiner, D.~Carmel, Q.~He, and J.~X. Yu, Eds.\hskip 1em plus
  0.5em minus 0.4em\relax {ACM}, 2019, pp. 5--6.

\end{thebibliography}

\begin{IEEEbiography}[{\includegraphics[width=1in,height=1.25in,clip,keepaspectratio]{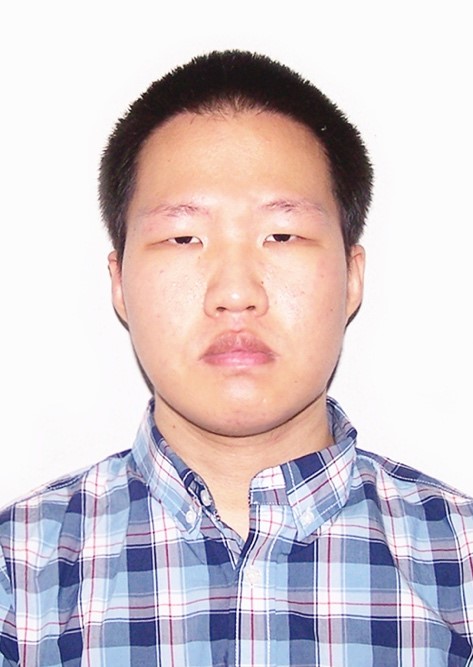}}]{Rui Kang} was in the School of Computer Science, Harbin Institute of Technology, Harbin, Heilongjiang, China while working on this paper. His research interests include integrity constraints and graph processing.
\end{IEEEbiography}

\begin{IEEEbiography}[{\includegraphics[width=1in,height=1.25in,clip,keepaspectratio]{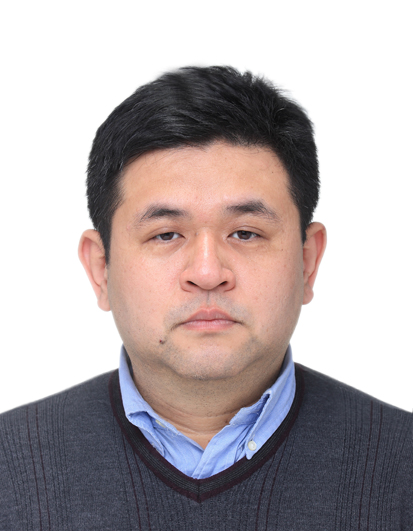}}]{Hongzhi Wang}
	Professor, PHD supervisor, the leader of massive data computing center and the vice dean of the honors school of Harbin Institute of Technology, the secretary general of ACM SIGMOD China, outstanding CCF member, a standing committee member CCF databases and a member of CCF big data committee. Research Fields include big data management and analysis, database systems, knowledge engineering and data quality. He was “starring track” visiting professor at MSRA and postdoctoral fellow at University of California, Irvine. Prof. Wang has been PI for more than 10 national or international projects including NSFC key project, NSFC projects and National Technical support project, and co-PI for more than 10 national projects include 973 project, 863 project and NSFC key projects. He also serves as a member of ACM Data Science Task Force. He has won First natural science prize of Heilongjiang Province, MOE technological First award, Microsoft Fellowship, IBM PHD Fellowship and Chinese excellent database engineer. His publications include over 200 papers in the journals and conferences such as VLDB Journal, IEEE TKDE, VLDB, SIGMOD, ICDE and SIGIR, 6 books and 3 book chapters. His PHD thesis was elected to be outstanding PHD dissertation of CCF and Harbin Institute of Technology. He severs as the reviewer of more than 20 international journal including VLDB Journal, IEEE TKDE, and PC members of over 50 international conferences including SIGMOD 2022, VLDB 2021, KDD 2021, ICML 2021, NeurpIS 2020, ICDE 2020, etc. His papers were cited more than 2000 times. His personal website is http://homepage.hit.edu.cn/wang.
\end{IEEEbiography}

\end{document}